\documentclass[useAMS,usenatbib,usegraphicx]{mn2e} \def\kms{km s$^{-1}$}
  \def\aap{A\&A} \def\apjl{ApJ}
\def\apj{ApJ} \def\apjs{ApJS} \def\aj{AJ} \def\mnras{MNRAS} \def\nat{Nature}
\def\pasp{PASP}  

\title[New Halo WD Candidates]
{New Halo White Dwarf Candidates in the Sloan Digital Sky Survey\thanks{Based
on observations obtained at the MMT Observatory, a joint facility of
the Smithsonian Institution and the University of Arizona.}}

\author[K. Dame et al.]
{Kyra Dame$^{1}$,
A. Gianninas$^{1}$,
Mukremin Kilic$^{1}$,
Jeffrey A. Munn$^{2}$,
Warren R. Brown$^{3}$,
\newauthor Kurtis A. Williams$^{4}$,
Ted von Hippel$^{5}$,
Hugh C. Harris$^{2}$\\
$^{1}$Homer L. Dodge Department of Physics \& Astronomy, University of Oklahoma,
440 W. Brooks St, Norman, OK 73019, USA\\
$^{2}$US Naval Observatory, Flagstaff Station, 10391 W. Naval Observatory Road,
Flagstaff, AZ 86005, USA\\
$^{3}$Smithsonian Astrophysical Observatory, 60 Garden St., Cambridge, MA 02138, USA\\
$^{4}$Department of Physics and Astronomy, Texas A\&M University-Commerce, P.O. Box 3011, Commerce, TX 75429, USA\\
$^{5}$Embry-Riddle Aeronautical University, Physical Sciences, 600 South Clyde Morris Boulevard Daytona Beach, FL 32114-3900, USA\\
}

\begin{document}

\maketitle

\begin{abstract}

We present optical spectroscopy and near-infrared photometry of 57 faint ($g= 19-22$) high proper motion white dwarfs
identified through repeat imaging of $\approx3100$ square degrees of the Sloan Digital Sky Survey footprint by
\citet{munn14}. We use $ugriz$ and $JH$ photometry to perform a model atmosphere analysis, and identify ten
ultracool white dwarfs with $T_{\rm eff}<4000$ K, including the coolest pure H atmosphere white dwarf currently
known, J1657+2638, with $T_{\rm eff}= 3550 \pm 100$K. The majority of the objects with cooling ages larger than
9 Gyr display thick disc kinematics and constrain the age of the thick disc to $\geq11$ Gyr.
There are four white dwarfs in our sample with large tangential
velocities ($v_{tan} > 120$ km $s^{-1}$) and UVW velocities that are more consistent with the halo than the
Galactic disc. For typical $0.6 M_{\odot}$ white dwarfs, the cooling ages for these halo candidates range from 2.3
to 8.5 Gyr. However, the total main-sequence + white dwarf cooling ages of these stars would be consistent
with the Galactic halo if they are slightly undermassive. 
Given the magnitude limits of the current large scale surveys, many of the coolest and oldest white dwarfs remain
undiscovered in the solar neighborhood, but upcoming surveys such as \textit{GAIA} and the Large Synoptic Survey
Telescope (LSST) should find many of these elusive thick disc and halo white dwarfs.
\end{abstract}


\section{Introduction}

As the remnants of some of the oldest stars in the galaxy, cool white dwarfs offer an
independent method for dating different Galactic populations and constraining their star
formation history \citep{winget87,liebert88}. The current best estimates for the ages of the
Galactic thin and thick discs are $8 \pm 1.5$ Gyr \citep{leggett98,harris06} and $\geq10$ Gyr
\citep{gianninas15}, respectively. Extended \textit{Hubble Space Telescope} observing campaigns
on 47 Tuc, M4, and NGC 6397 revealed the end of the white dwarf cooling sequence in these globular
clusters \citep{hansen04,hansen07,kalirai12b}, which reveal an age spread of 11 to 13 Gyr for the
Galactic halo \citep{campos15}. 

Field white dwarfs provide additional and superior information on the age and age spread of the
Galactic disc and halo. Recent large scale surveys such as the Sloan Digital Sky Survey (SDSS) have found many cool field
white dwarfs \citep{gates04,harris06,harris08,kilic06,kilic10b,vidrih07,tremblay14,gianninas15}. These stars are
far closer and brighter than those found in globular clusters, allowing for relatively easy optical and infrared
observations in multiple bands from ground-based telescopes. Modeling the spectral energy distributions (SEDs)
of these white dwarfs provides excellent constraints on their atmospheric composition and cooling ages and
gives us an alternate method for calibrating the white dwarf cooling sequences of globular clusters.
However, there are only a handful of nearby halo white dwarfs currently known.

\citet{kalirai12a} use four field white dwarfs with halo kinematics to derive an age of $11.4 \pm 0.7$ Gyr for the
inner halo. These four stars are warm enough to show Balmer absorption lines, which enable precise constraints
on their surface gravity, mass, temperature, and cooling ages. The SPY project (ESO SN Ia Progenitor 
Survey) found 12 halo members with ages consistent with a halo age $\approx 11$ Gyr in a sample of 634 DA white dwarfs. These
have accurate radial velocities determined from Balmer absorption lines, allowing for the determination of accurate 3D space 
velocities \citep{pauli03,pauli06,richter07}. Similarly, \citet{kilic12} use optical and infrared photometry and parallax 
observations of two cool white dwarfs with halo kinematics, WD 0346+246 and SDSS J110217.48+411315.4 to derive an age of 
11.0-11.5 Gyr for the local halo. 

Ongoing and future photometric and astrometric surveys like the Panoramic Survey Telescope \& Rapid Response System
\citep{tonry12}, Palomar Transient Factory \citep{rau09}, the Large Synoptic
Survey Telescope and the \textit{GAIA} mission will significantly increase the number of field white dwarfs known.
Previously, \citet{liebert07} performed a targeted proper motion survey for identifying thick disc and halo
white dwarfs in the solar neighborhood. \citet{munn14} present the proper motion catalog from this survey,
which includes $\approx 3100$ square degrees of sky observed at the Steward Observatory Bok 90 inch telescope
and the U.S. Naval Observatory Flagstaff Station 1.3 m telescope. \citet{kilic10a} presented three halo
white dwarf candidates identified in this survey, with ages of 10-11 Gyr. Here we present follow-up observations
and model atmosphere analysis of 54 additional high proper motion white dwarfs identified in this survey.
We find seven new ultracool ($T_{\rm eff}<4000$ K) white dwarfs and three new halo white dwarf candidates.
We discuss the details of our observations in Section \ref{sec:obs}, and the model atmosphere fits in Section
\ref{sec:phot}. We present the kinematic analysis of our sample in Section \ref{sec:kinematics}, and conclude
in Section \ref{sec:con}.

\section{Target Selection and Observations}
\label{sec:obs}

\subsection{The Reduced Proper Motion Diagram}

\begin{figure}
 \includegraphics[width = 9cm]{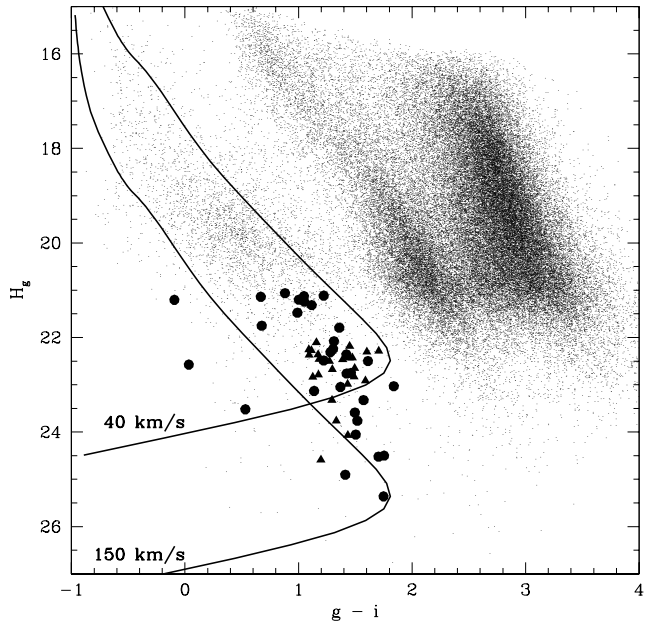}
 \caption{The reduced proper motion diagram for a portion of the \citet{munn14} proper motion
survey. White dwarf evolutionary tracks for tangential velocities of 40 and 150 \kms\ are shown as solid
lines. Filled circles mark spectroscopically confirmed white dwarfs with $H_g>21$ mag and triangles show
our targets with SWIRC near-infrared photometry, but with no follow-up spectroscopy.}
 \label{fig:rpm}
\end{figure}

Reduced proper motion is defined as $H = m + 5 \log{\mu} + 5$ (where $\mu$ is the proper motion and m is the apparent magnitude), which is equivalent to $M + 5 \log{V_{\rm tan}} - 3.379$.
Hence, reduced proper motion can be used to identify samples with similar kinematics, like the disc or halo white
dwarf population. \citet{kilic06,kilic10b} demonstrate that the reduced proper motion diagram provides a clean
sample of white dwarfs, with contamination rates of 1\% from halo subdwarfs. Figure \ref{fig:rpm} displays the
reduced proper motion diagram for a portion of the sky covered by the \citet{munn14} proper motion survey.
Going from left to right, three distinct populations of objects are clearly visible in this diagram;
white dwarfs, halo subdwarfs, and disc dwarfs. The solid lines show the predicted evolutionary sequences
for $\log{g}=8$ white dwarfs with $V_{\rm tan}=40$ and 150 \kms. The model colors become redder until the white
dwarfs become cool enough to show infrared absorption due to molecular hydrogen \citep{hansen98}.
We selected targets for follow-up spectroscopy and near-infrared photometry based on their reduced proper motion
and colors. To find the elusive halo white dwarfs and other white dwarfs with high tangential velocities, we
targeted objects with $H_g>21$ mag and below the $V_{\rm tan}=40$ \kms\ line.

\subsection{Optical Spectroscopy}

We obtained follow-up optical spectroscopy of 32 white dwarf candidates at the 6.5 m MMT telescope
equipped with the Blue Channel Spectrograph \citep{schmidt89} on UT 2009 June 18-23 and 2009 November 19-20. 
We used a $1\farcs25$ slit and the 500 line mm$^{-1}$ grating in first order to obtain
spectra over the range 3660-6800 \AA{} and with a resolving power of R = 1200. 
We obtained all spectra at the parallactic angle and acquired He-Ar-Ne comparison lamp
exposures for wavelength calibration. We use observations of the cool white dwarf G24-9
for flux calibration. 

Out of the 32 candidates with spectra, only two (SDSS J024416.07-090919.7 and
J172431.61+261543.1) are metal-poor halo subdwarfs. The remaining objects are confirmed to
be DA, DC, or DZ white dwarfs. This relatively small (2 out of 32) contamination rate from 
subdwarfs demonstrates that our white dwarf sample is relatively clean.

\begin{figure}
 \includegraphics[width = 9cm]{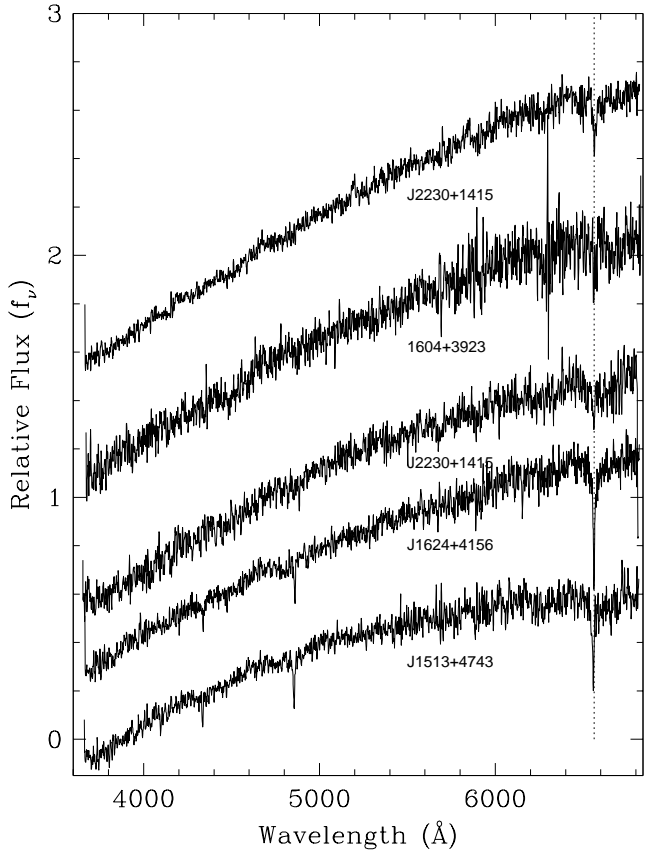}
 \caption{Optical spectra for the five DA WDs in our sample. The dotted line marks H$\alpha$.}
 \label{fig:DA}
\end{figure}

Figure \ref{fig:DA} shows the spectra for the five DA WDs in our sample. Two of the DAs,
J1513+4743 and J1624+4156, are warm enough ($T_{\rm eff}\approx 5900$ K) to show H$\alpha$
and a few of the higher order Balmer lines, while the remaining three DAs only show H$\alpha$,
which implies effective temperatures near 5000 K.

\begin{figure}
 \includegraphics[width = 9cm]{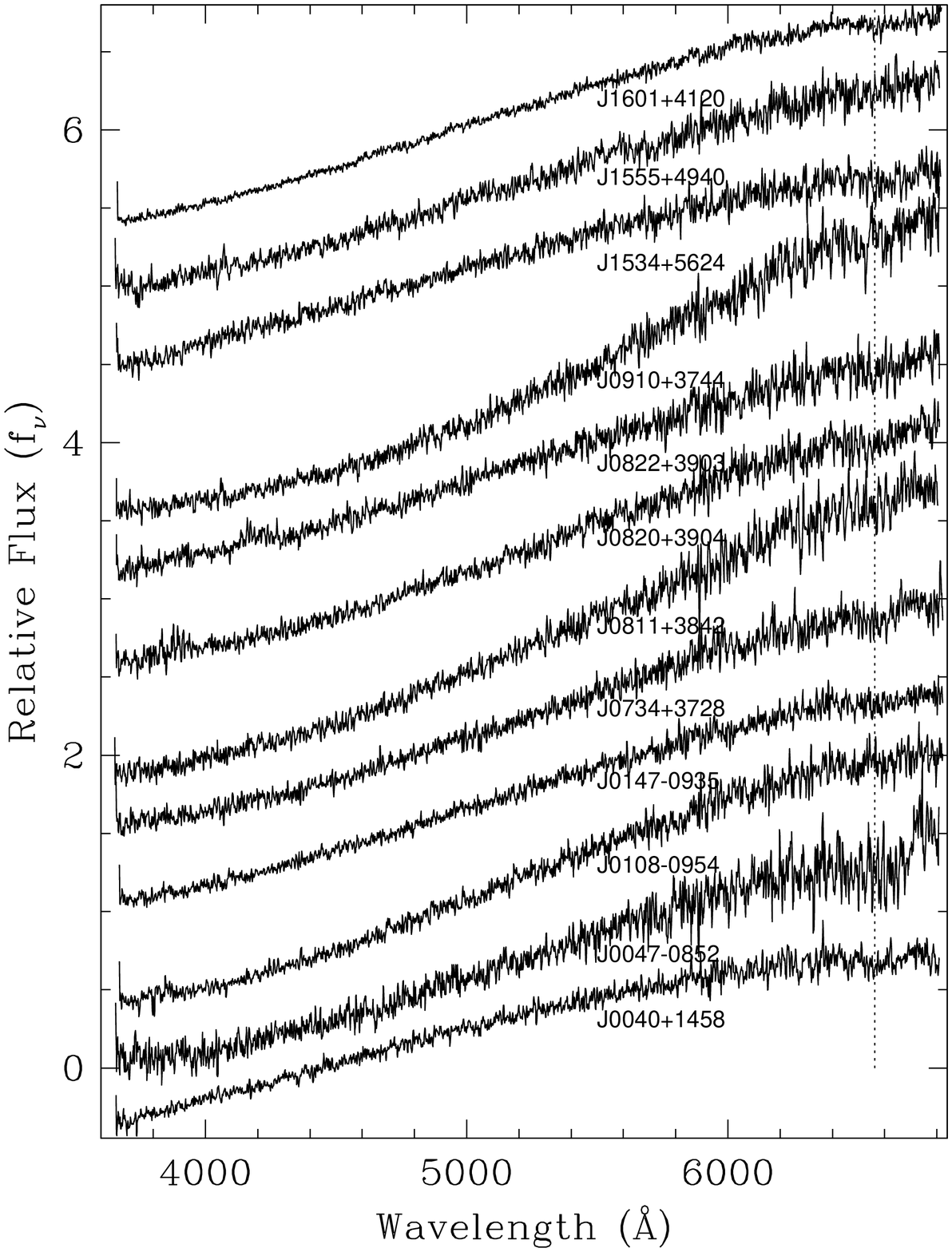}
 \includegraphics[width = 9cm]{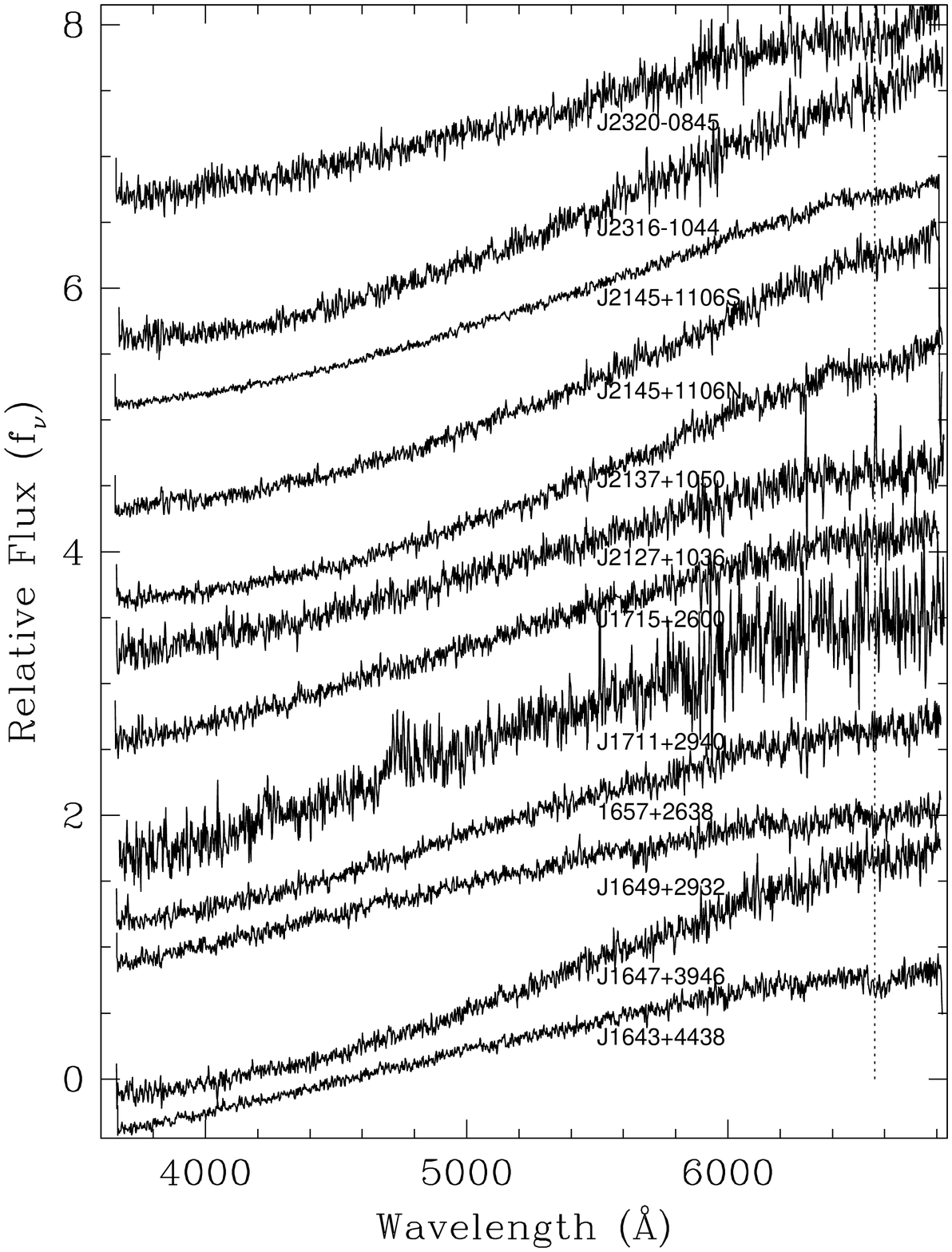}
 \caption{Optical spectra for 24 DC WDs in our sample.}
 \label{fig:DC}
\end{figure}

Figure \ref{fig:DC} shows the MMT spectra for the 24 DC white dwarfs in our sample, including
the three cool DCs presented in \citet{kilic10a}. All of these 24 targets have featureless
spectra that are rising toward the infrared, indicating temperatures below 5000 K.  

\subsection{Near-Infrared Photometry}

We obtained J- and H-band imaging observations of 40 of our targets using the Smithsonian Widefield Infrared Camera
\citep[SWIRC,][]{brown08} on the MMT
on UT 2011 March 23-24. SWIRC has a $5.12 \times 5.12$ arcmin field of view at a resolution of $0\farcs15$ per pixel.
We observed each target on a dozen or more dither positions, and obtained dark frames and
sky flats each evening. We used the SWIRC data reduction pipeline to perform
dark correction, flat-fielding, and sky subtraction, and to produce a combined image for each field in each filter.
We use the 2MASS stars in the SWIRC field of view for photometric  and astrometric calibration.
In addition, near-infrared photometry for two more targets, J0040+1458 and J1649+2932, are available from the
UKIRT Infrared Deep Sky Survey \citep[UKIDSS,][]{lawrence07} Large Area Survey. 
Table \ref{tab:phot} presents the \textit{ugriz} and \textit{JH} photometry for our sample
of 57 targets with follow-up spectroscopy and/or near-infrared photometry.

\begin{figure}
\includegraphics[angle=-90,width = 9cm]{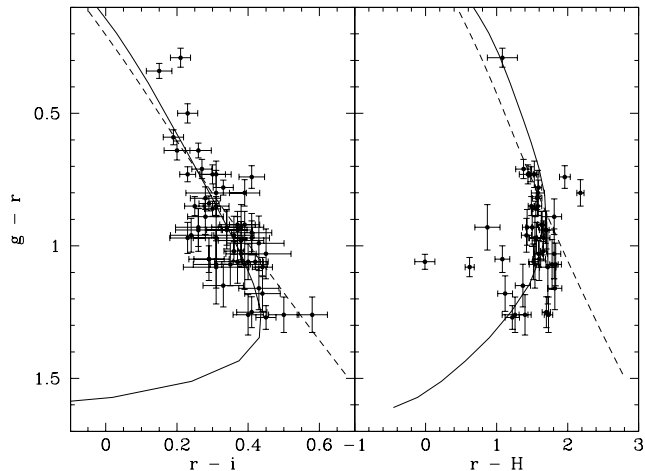}
\caption{Color-color diagrams for our sample of 57 white dwarf candidates.
Solid and dashed lines show the predicted colors for pure H ($T_{\rm eff} \geq 2500$ K)
and pure He atmosphere ($T_{\rm eff} \geq 3500$ K) white dwarfs with $\log{g}= 8$
\citep{bergeron95}, respectively.}
 \label{fig:color}
\end{figure}

Figure \ref{fig:color} presents optical and infrared color-color diagrams for the same stars,
along with the predicted colors of pure H and pure He atmosphere white dwarfs. The differences
between these models are relatively minor in the optical color-color diagrams, except for the
ultracool white dwarfs with $T_{\rm eff}<4000$ K. The pure H
models predict the colors to get redder until the onset of the Collision Induced Absorption (CIA)
due to molecular hydrogen, which leads to a blue hook feature. This transition occurs at 3750 K
for the $r-i$ color, whereas it occurs at 4500 K for the $r-H$ color. The colors for our
sample of 57 stars, including the targets with and without follow-up spectroscopy, are consistent
with the white dwarf model colors within the errors. The majority of the targets with $g-r\geq1.0$ mag
($T_{\rm eff}\leq4250$ K) show bluer $r-H$ colors than the pure He model sequence, indicating
that the coolest white dwarfs in our sample have H-rich atmospheres.

We use the SWIRC astrometry to verify the proper motion measurements from our optical
imaging survey. Given the relatively small field of view of the SWIRC camera and the limited
number of 2MASS stars available in each field, the astrometric precision is significantly worse
in the SWIRC images compared to the Bok 90 inch and USNO 1.3m optical data. We find that the proper
motion measurements from the SWIRC data are on average $54\pm44$ mas yr$^{-1}$ higher. Nevertheless,
all but one of our targets, J1513+4743, have SWIRC-SDSS proper motion measurements consistent with the
proper motion measurements from our optical data within 3$\sigma$. J1513+4743 is spectroscopically
confirmed to be a DA white dwarf. Hence, the contamination rate of our sample of 57 stars by objects
with incorrectly measured proper motions should be relatively small.  

\begin{table*}
\centering
\scriptsize
\caption{Optical and Near-Infrared Photometry of White Dwarf Candidates}
\label{tab:phot}
\begin{tabular}{lccccccc}
\hline
SDSS               &        \textit{u}         &        \textit{g}         &        \textit{r}         &        \textit{i}         &        \textit{z}         &       \textit{J}          &        \textit{H}        \\ \hline
J213730.86+105041.5  & 23.30 $\pm$ 0.54 & 21.77 $\pm$ 0.06 & 20.51 $\pm$ 0.03 & 20.01 $\pm$ 0.03 & 19.73 $\pm$ 0.08 & 19.21 $\pm$ 0.10 & 19.25 $\pm$ 0.18 \\
J214538.16+110626.6  & 23.72 $\pm$ 0.52 & 21.49 $\pm$ 0.04 & 20.22 $\pm$ 0.02 & 19.77 $\pm$ 0.02 & 19.61 $\pm$ 0.05 & 18.87 $\pm$ 0.07 & 19.00 $\pm$ 0.10 \\
J214538.60+110619.1  & 23.47 $\pm$ 0.45 & 21.01 $\pm$ 0.03 & 19.93 $\pm$ 0.02 & 19.49 $\pm$ 0.02 & 19.29 $\pm$ 0.04 & 18.54 $\pm$ 0.06 & 19.31 $\pm$ 0.06 \\
J004022.47+145835.0 & 22.23 $\pm$ 0.23 & 20.56 $\pm$ 0.03 & 19.83 $\pm$ 0.02 & 19.53 $\pm$ 0.03 & 19.43 $\pm$ 0.06 & 18.60 $\pm$ 0.10 & 18.39 $\pm$ 0.12 \\
J004725.61$-$085223.9 & 25.78 $\pm$ 0.74 & 21.65 $\pm$ 0.06 & 20.67 $\pm$ 0.04 & 20.29 $\pm$ 0.05 & 20.17 $\pm$ 0.15 & \dots              & \dots              \\
J010838.42$-$095415.7 & 24.07 $\pm$ 1.04 & 21.56 $\pm$ 0.06 & 20.61 $\pm$ 0.04 & 20.20 $\pm$ 0.04 & 19.98 $\pm$ 0.13 & \dots & \dots                       \\
J014749.07$-$093537.4 & 22.98 $\pm$ 0.53 & 21.64 $\pm$ 0.07 & 20.70 $\pm$ 0.04 & 20.44 $\pm$ 0.05 & 20.39 $\pm$ 0.24 & \dots & \dots                       \\
J073417.76+372842.6 & 24.01 $\pm$ 0.74 & 21.88 $\pm$ 0.07 & 20.95 $\pm$ 0.05 & 20.57 $\pm$ 0.04 & 20.44 $\pm$ 0.14 & 19.59 $\pm$ 0.09 & 20.08 $\pm$ 0.17 \\
J074942.95+294716.7 & 22.45 $\pm$ 0.26 & 20.80 $\pm$ 0.03 & 19.95 $\pm$ 0.02 & 19.64 $\pm$ 0.03 & 19.49 $\pm$ 0.06 & 18.56 $\pm$ 0.04 & 18.37 $\pm$ 0.05 \\
J080505.26+273557.2 & 23.73 $\pm$ 0.62 & 21.52 $\pm$ 0.06 & 20.59 $\pm$ 0.03 & 20.25 $\pm$ 0.03 & 20.28 $\pm$ 0.12 & 19.05 $\pm$ 0.06 & 19.16 $\pm$ 0.06 \\
J080545.80+374720.4 & 23.87 $\pm$ 0.64 & 21.12 $\pm$ 0.04 & 20.39 $\pm$ 0.03 & 20.08 $\pm$ 0.03 & 19.87 $\pm$ 0.10 & 19.13 $\pm$ 0.05 & 18.86 $\pm$ 0.06 \\
J081140.07+384202.2 & 23.87 $\pm$ 0.69 & 21.89 $\pm$ 0.07 & 20.84 $\pm$ 0.04 & 20.55 $\pm$ 0.04 & 20.30 $\pm$ 0.11 & 19.50 $\pm$ 0.06 & 19.24 $\pm$ 0.08 \\
J081735.51+310625.5 & 22.39 $\pm$ 0.23 & 20.52 $\pm$ 0.03 & 19.50 $\pm$ 0.02 & 19.14 $\pm$ 0.02 & 18.89 $\pm$ 0.03 & 18.14 $\pm$ 0.03 & 17.84 $\pm$ 0.04 \\
J082035.23+390419.9 & 22.72 $\pm$ 0.32 & 22.01 $\pm$ 0.07 & 20.86 $\pm$ 0.04 & 20.53 $\pm$ 0.04 & 20.29 $\pm$ 0.09 & 19.31 $\pm$ 0.05 & 19.49 $\pm$ 0.10 \\
J082255.41+390302.7 & 23.16 $\pm$ 0.34 & 21.87 $\pm$ 0.05 & 20.90 $\pm$ 0.04 & 20.49 $\pm$ 0.03 & 20.37 $\pm$ 0.11 & 19.19 $\pm$ 0.06 & 19.24 $\pm$ 0.08 \\
J082842.31+352729.5 & 21.43 $\pm$ 0.09 & 19.84 $\pm$ 0.02 & 19.06 $\pm$ 0.02 & 18.73 $\pm$ 0.02 & 18.69 $\pm$ 0.03 & 17.88 $\pm$ 0.03 & 17.47 $\pm$ 0.04 \\
J084802.30+420429.7 & 22.92 $\pm$ 0.37 & 21.72 $\pm$ 0.06 & 20.79 $\pm$ 0.04 & 20.53 $\pm$ 0.05 & 20.41 $\pm$ 0.14 & 19.63 $\pm$ 0.06 & 19.29 $\pm$ 0.08 \\
J085441.14+390700.1 & 22.87 $\pm$ 0.28 & 21.22 $\pm$ 0.03 & 20.40 $\pm$ 0.03 & 20.12 $\pm$ 0.03 & 20.03 $\pm$ 0.08 & 19.17 $\pm$ 0.04 & 18.83 $\pm$ 0.06 \\
J091035.82+374454.8 & 23.85 $\pm$ 0.63 & 21.79 $\pm$ 0.06 & 20.53 $\pm$ 0.03 & 19.95 $\pm$ 0.03 & 19.64 $\pm$ 0.07 & 18.95 $\pm$ 0.04 & 18.80 $\pm$ 0.05 \\
J091823.08+502826.4 & 22.60 $\pm$ 0.26 & 20.72 $\pm$ 0.03 & 19.87 $\pm$ 0.02 & 19.62 $\pm$ 0.02 & 19.47 $\pm$ 0.06 & 18.67 $\pm$ 0.04 & 18.38 $\pm$ 0.04 \\
J092716.99+485233.3 & 22.96 $\pm$ 0.30 & 20.65 $\pm$ 0.02 & 19.59 $\pm$ 0.02 & 19.17 $\pm$ 0.03 & 19.11 $\pm$ 0.05 & 19.12 $\pm$ 0.06 & 19.60 $\pm$ 0.14 \\
J100953.03+534732.9 & 23.82 $\pm$ 0.84 & 21.82 $\pm$ 0.07 & 20.75 $\pm$ 0.04 & 20.40 $\pm$ 0.04 & 19.86 $\pm$ 0.10 & 19.09 $\pm$ 0.06 & 18.92 $\pm$ 0.07 \\
J102417.17+492011.3 & 22.83 $\pm$ 0.29 & 21.59 $\pm$ 0.06 & 20.70 $\pm$ 0.03 & 20.42 $\pm$ 0.04 & 20.15 $\pm$ 0.10 & 19.46 $\pm$ 0.08 & 18.89 $\pm$ 0.10 \\
J105652.84+504321.3 & 23.56 $\pm$ 0.56 & 21.40 $\pm$ 0.04 & 20.37 $\pm$ 0.03 & 19.99 $\pm$ 0.03 & 19.79 $\pm$ 0.08 & 19.05 $\pm$ 0.04 & 18.78 $\pm$ 0.05 \\
J110105.01+485437.9 & 23.00 $\pm$ 0.51 & 20.88 $\pm$ 0.04 & 19.92 $\pm$ 0.05 & 19.68 $\pm$ 0.03 & 19.67 $\pm$ 0.09 & 18.75 $\pm$ 0.04 & 18.50 $\pm$ 0.05 \\
J114558.52+563806.8 & 23.08 $\pm$ 0.52 & 21.73 $\pm$ 0.06 & 20.93 $\pm$ 0.06 & 20.62 $\pm$ 0.06 & 20.26 $\pm$ 0.15 & 19.64 $\pm$ 0.08 & 19.35 $\pm$ 0.07 \\
J120514.49+550217.2 & 22.84 $\pm$ 0.43 & 21.31 $\pm$ 0.04 & 20.37 $\pm$ 0.03 & 20.00 $\pm$ 0.03 & 19.88 $\pm$ 0.10 & 18.94 $\pm$ 0.05 & 18.68 $\pm$ 0.04 \\
J132358.81+022342.2 & 23.10 $\pm$ 0.46 & 21.88 $\pm$ 0.07 & 20.91 $\pm$ 0.04 & 20.60 $\pm$ 0.05 & 20.52 $\pm$ 0.16 & 19.43 $\pm$ 0.07 & 19.37 $\pm$ 0.07 \\
J133309.98+494227.2 & 24.03 $\pm$ 0.78 & 21.15 $\pm$ 0.04 & 20.23 $\pm$ 0.03 & 19.86 $\pm$ 0.04 & 19.60 $\pm$ 0.06 & 18.79 $\pm$ 0.04 & 18.60 $\pm$ 0.04 \\
J140907.89$-$010036.9 & 22.94 $\pm$ 0.28 & 21.64 $\pm$ 0.06 & 20.58 $\pm$ 0.03 & 20.18 $\pm$ 0.03 & 19.97 $\pm$ 0.10 & 19.12 $\pm$ 0.06 & 19.06 $\pm$ 0.06 \\
J142136.69+035612.4 & 24.39 $\pm$ 0.97 & 21.89 $\pm$ 0.08 & 20.82 $\pm$ 0.05 & 20.43 $\pm$ 0.04 & 20.23 $\pm$ 0.16 & 19.19 $\pm$ 0.06 & 19.04 $\pm$ 0.07 \\
J143400.55+534525.2 & 22.95 $\pm$ 0.44 & 21.20 $\pm$ 0.04 & 20.28 $\pm$ 0.03 & 19.89 $\pm$ 0.03 & 19.62 $\pm$ 0.07 & 18.99 $\pm$ 0.06 & 18.64 $\pm$ 0.06 \\
J144417.48+602555.1 & 23.73 $\pm$ 0.61 & 21.62 $\pm$ 0.05 & 20.65 $\pm$ 0.03 & 20.42 $\pm$ 0.04 & 20.52 $\pm$ 0.13 & 19.50 $\pm$ 0.07 & 19.09 $\pm$ 0.10 \\
J144606.46+025811.5 & 23.70 $\pm$ 0.85 & 21.72 $\pm$ 0.12 & 20.64 $\pm$ 0.07 & 20.33 $\pm$ 0.06 & 30.02 $\pm$ 0.12 & 19.03 $\pm$ 0.06 & 18.92 $\pm$ 0.09 \\
J150904.50+540825.2 & 24.60 $\pm$ 1.03 & 21.97 $\pm$ 0.08 & 20.94 $\pm$ 0.05 & 20.49 $\pm$ 0.05 & 20.17 $\pm$ 0.13 & 19.31 $\pm$ 0.07 & 19.13 $\pm$ 0.08 \\
J151319.26+502318.6 & 23.39 $\pm$ 0.44 & 21.96 $\pm$ 0.07 & 20.70 $\pm$ 0.03 & 20.30 $\pm$ 0.03 & 19.93 $\pm$ 0.07 & 18.85 $\pm$ 0.05 & 19.30 $\pm$ 0.08 \\
J151321.20+474324.2 & 20.82 $\pm$ 0.06 & 19.92 $\pm$ 0.03 & 19.63 $\pm$ 0.02 & 19.42 $\pm$ 0.02 & 19.41 $\pm$ 0.06 & 18.62 $\pm$ 0.05 & 18.55 $\pm$ 0.21 \\
J151555.53+593045.3 & 25.19 $\pm$ 0.86 & 21.96 $\pm$ 0.06 & 20.78 $\pm$ 0.03 & 20.34 $\pm$ 0.04 & 20.24 $\pm$ 0.09 & 19.40 $\pm$ 0.08 & 19.66 $\pm$ 0.10 \\
J153300.94$-$001212.2 & 23.65 $\pm$ 0.59 & 22.05 $\pm$ 0.07 & 20.89 $\pm$ 0.04 & 20.46 $\pm$ 0.04 & 20.16 $\pm$ 0.12 & 19.24 $\pm$ 0.07 & 19.07 $\pm$ 0.09 \\
J153432.25+562455.7 & 21.76 $\pm$ 0.14 & 20.26 $\pm$ 0.02 & 19.62 $\pm$ 0.02 & 19.36 $\pm$ 0.03 & 19.28 $\pm$ 0.05 & \dots & \dots                       \\
J155243.40+463819.4 & 21.53 $\pm$ 0.11 & 20.09 $\pm$ 0.02 & 19.50 $\pm$ 0.02 & 19.31 $\pm$ 0.02 & 19.28 $\pm$ 0.05 & \dots & \dots                       \\
J155501.57+494056.4 & 25.76 $\pm$ 0.64 & 21.84 $\pm$ 0.08 & 20.77 $\pm$ 0.04 & 20.46 $\pm$ 0.05 & 20.30 $\pm$ 0.16 & 19.29 $\pm$ 0.07 & 19.23 $\pm$ 0.07 \\
J160125.48+412014.1 & 21.20 $\pm$ 0.07 & 19.28 $\pm$ 0.02 & 18.44 $\pm$ 0.02 & 18.15 $\pm$ 0.02 & 18.08 $\pm$ 0.02 & \dots & \dots                       \\
J160130.82+420427.6 & 24.19 $\pm$ 0.98 & 21.65 $\pm$ 0.06 & 20.71 $\pm$ 0.04 & 20.37 $\pm$ 0.04 & 20.37 $\pm$ 0.17 & 19.29 $\pm$ 0.07 & 19.01 $\pm$ 0.09 \\
J160424.38+392330.5 & 21.88 $\pm$ 0.20 & 20.51 $\pm$ 0.03 & 19.87 $\pm$ 0.02 & 19.67 $\pm$ 0.03 & 19.44 $\pm$ 0.07 & \dots & \dots                       \\
J162417.93+415656.6 & 21.21 $\pm$ 0.09 & 20.18 $\pm$ 0.02 & 19.84 $\pm$ 0.02 & 19.69 $\pm$ 0.03 & 19.62 $\pm$ 0.07 & \dots & \dots                       \\
J162724.57+372643.1 & 21.78 $\pm$ 0.10 & 19.80 $\pm$ 0.02 & 18.94 $\pm$ 0.02 & 18.64 $\pm$ 0.01 & 18.53 $\pm$ 0.03 & 17.60 $\pm$ 0.03 & 17.39 $\pm$ 0.03 \\
J164358.79+443855.4 & 21.42 $\pm$ 0.10 & 19.84 $\pm$ 0.02 & 19.11 $\pm$ 0.02 & 18.88 $\pm$ 0.01 & 18.81 $\pm$ 0.04 & 17.91 $\pm$ 0.03 & 17.63 $\pm$ 0.03 \\
J164745.45+394638.6 & 24.56 $\pm$ 1.09 & 21.55 $\pm$ 0.05 & 20.30 $\pm$ 0.03 & 19.89 $\pm$ 0.03 & 19.84 $\pm$ 0.08 & 18.91 $\pm$ 0.05 & 18.60 $\pm$ 0.05 \\
J164931.91+293247.7 & 22.19 $\pm$ 0.15 & 20.61 $\pm$ 0.03 & 19.90 $\pm$ 0.02 & 19.63 $\pm$ 0.02 & 19.51 $\pm$ 0.08 & 18.62 $\pm$ 0.06 & 18.52 $\pm$ 0.11 \\
J165723.84+263843.5 & 24.80 $\pm$ 0.80 & 21.33 $\pm$ 0.04 & 20.28 $\pm$ 0.03 & 19.99 $\pm$ 0.03 & 19.77 $\pm$ 0.10 & 19.24 $\pm$ 0.08 & 19.20 $\pm$ 0.10 \\
J171135.27+294046.0 & 23.00 $\pm$ 0.33 & 21.11 $\pm$ 0.03 & 20.37 $\pm$ 0.03 & 19.96 $\pm$ 0.02 & 19.85 $\pm$ 0.07 & 18.73 $\pm$ 0.05 & 18.41 $\pm$ 0.07\\
J171543.76+260016.9 & 22.82 $\pm$ 0.40 & 21.25 $\pm$ 0.04 & 20.45 $\pm$ 0.03 & 20.06 $\pm$ 0.03 & 19.91 $\pm$ 0.10 & 18.81 $\pm$ 0.07 & 18.27 $\pm$ 0.04 \\
J212739.00+103655.1 & 22.77 $\pm$ 0.41 & 21.67 $\pm$ 0.06 & 20.71 $\pm$ 0.04 & 20.35 $\pm$ 0.04 & 20.11 $\pm$ 0.13 & \dots & \dots                       \\
J223038.21+141505.7 & 21.86 $\pm$ 0.15 & 20.37 $\pm$ 0.03 & 19.87 $\pm$ 0.02 & 19.64 $\pm$ 0.02 & 19.60 $\pm$ 0.07 & \dots & \dots                       \\
J231617.67$-$104411.0 & 23.41 $\pm$ 0.67 & 21.98 $\pm$ 0.09 & 20.99 $\pm$ 0.05 & 20.56 $\pm$ 0.05 & 20.34 $\pm$ 0.14 & \dots & \dots                       \\
J232018.23$-$084516.7 & 25.55 $\pm$ 0.95 & 21.63 $\pm$ 0.07 & 20.57 $\pm$ 0.04 & 20.20 $\pm$ 0.05 & 19.97 $\pm$ 0.15 & \dots & \dots\\
\hline
\end{tabular}
\end{table*}

\section{Photometric Analysis}
\label{sec:phot}

\subsection{Model Atmospheres}

Our model atmospheres come from the LTE model atmosphere code described in \citet{bergeron95}
and references within, along with the recent improvements in the calculations for the Stark
broadening of hydrogen lines discussed in \citet{tremblay09}. We follow the method of \citet{holberg06} and
convert the observed magnitudes into fluxes, and use a nonlinear least-squares method to fit the resulting
SEDs to predictions from model atmospheres. Given that all our targets appear to be within 150 pc, we do not 
correct for extinction. We consider only the temperature and the solid angle $\pi(R/D)^{2}$, where $R$ is the 
radius of the white dwarf and $D$ is its distance from Earth, as free parameters. Convection is modeled by the 
ML/$\alpha$ = 0.7 prescription of mixing length theory. For a more detailed discussion of our fitting technique, 
see \citet{bergeron01b}; for details of our helium-atmosphere models, see \citet{bergeron11}. Since we do not have 
parallax measurements for our objects, we assume a surface gravity of $\log{g}=8$. This is appropriate, as the 
white dwarf mass distribution in the Solar neighborhood peaks at about 0.6 $M_{\odot}$ \citep{tremblay13}.  We 
discuss the effects of this choice in Section \ref{sec:kinematics}.

Below about 5000 K, H$\alpha$ is not visible. However, the presence of hydrogen can still be seen in
the blue from the red wing of Ly$\alpha$ absorption \citep{kowalski06}, and in the infrared from CIA due to molecular hydrogen.
Cool white dwarfs with pure helium atmospheres are not subject to these opacities, so their SEDs should
appear similar to a blackbody. Because of this, atmospheric composition can still be determined from 
ultraviolet and near-infrared data. Table \ref{tab:wd} presents the best-fit atmospheric compositions,
temperatures, distances, and cooling ages for our targets, as well as their proper motions and tangential velocities. 
Below, we discuss the pure H, pure He, and mixed H/He atmosphere targets separately, and highlight the most interesting
objects in the sample. 

\begin{table*}
\centering
\scriptsize
\caption{Physical Parameters of our White Dwarf Sample. Source of the optical spectroscopic observations: (1) This paper, (2) \citet{kilic10b}, and (3) \citet{kilic10a}.}
\label{tab:wd}
\begin{tabular}{lccclrrrrr}
\hline
Object  & Spectral & Source & Composition & $T_{\rm eff}$ & $d$  & Cooling Age  & $\mu_{RA}$     & $\mu_{Dec}$    & $V_{tan}$ \\
(SDSS) &  Type  &  & (log He/H)  &   (K)               &     (pc)     &  (Gyr)  & (mas yr$^{-1}$)  & (mas yr$^{-1}$)    &   (\kms)  \\ 
\hline
J2137+1050   & DC   & 2 & H           & 3670 $\pm$ 160   & 75            & 9.8        & $-$228.9 & $-$473.6 & 187.1       \\
J2145+1106N  & DC   & 2 & H           & 3720 $\pm$ 110   & 68            & 9.7        & 191.9  & $-$366.9 & 134.4       \\
J2145+1106S  & DC   & 2 & H           & 3960 $\pm$ 100   & 65            & 9.1        & 185.9  & $-$367.7 & 126.5       \\                    
J0040+1458   & DC   & 1 & He          & 4890 $\pm$ 90    & 94            & 6.3        & 128.1  & 18.5   & 57.6        \\
J0047$-$0852 & DC   & 1 & H           & 3140 $\pm$ 160   & 68            & 11.0       & 211    & $-$27.2  & 69.0        \\
             &      &   & He          & 3920 $\pm$ 120   & 79            & 8.5        &        &        & 79.2        \\
J0108$-$0954 & DC   & 1 & H           & 3630 $\pm$ 520   & 77            & 9.9        & $-$70.1  & $-$183.9 & 72.1        \\
             &      &   & He          & 4360 $\pm$ 150   & 100           & 7.5        &        &        & 93.6        \\      
J0147$-$0935 & DC   & 1 & H           & 4300 $\pm$ 320   & 108           & 8.2        & 211.5  & $-$26.4  & 109.0       \\
             &      &   & He          & 4640 $\pm$ 190   & 126           & 6.9        &        &        & 127.3       \\                    
J0734+3728   & DC   & 1 & H           & 3700 $\pm$ 140   & 94            & 9.8        & $-$3.5   & $-$114   & 50.5        \\
J0749+2947   &\dots & \dots & He      & 4690 $\pm$ 60    & 90            & 6.7        & 216.1  & $-$133.9 & 108.9       \\
J0805+2735   &\dots & \dots & H       & 4130 $\pm$ 120   & 94            & 8.7        & 68.1   & $-$215.9 & 101.1       \\
J0805+3747   &\dots & \dots & He      & 4820 $\pm$ 80    & 118           & 6.4        & $-$91.8  & $-$135.5 & 91.4        \\
J0811+3842   & DC   & 1 &  He         & 4570 $\pm$ 110   & 131           & 7.0        & 99.3   & $-$147.4 & 110.1       \\
J0817+3106   &\dots & \dots & He      & 4510 $\pm$ 50    & 67            & 7.1        & 231.4  & $-$93.8  & 79.8        \\
J0820+3904   & DC   & 1 & H           & 4050 $\pm$ 150   & 104           & 8.9        & $-$172   & $-$129   & 106.0       \\
J0822+3903   & DC   & 1 & H           & 4190 $\pm$ 150   & 109           & 8.5        & 274.2  & $-$316.5 & 216.5       \\
J0828+3527   & DC   & 3 & He          & 4840 $\pm$ 60    & 65            & 6.4        & $-$13.1  & $-$161   & 49.9        \\
J0848+4204   &\dots & \dots & He      & 4820 $\pm$ 120   & 146           & 6.4        & $-$137.8 & $-$32.5  & 97.9        \\
J0854+3907   &\dots & \dots & He      & 4810 $\pm$ 80    & 119           & 6.5        & $-$29.9  & $-$162.1 & 93.2        \\
J0910+3744   & DC   & 1 & $-$3.69       & 3450 $\pm$ 190   & 63            & 9.5        & $-$143.6 & $-$91.7  & 50.7        \\
J0918+5028   &\dots & \dots & He      & 4810 $\pm$ 70    & 94            & 6.5        & $-$108.2 & $-$185.9 & 96.3        \\
J0927+4852   &\dots & \dots & 6.33    & 3210 $\pm$ 90    & 42            & 9.9        & 216.5  & $-$70.1  & 45.4        \\
J1009+5347   &\dots & \dots & He      & 4290 $\pm$ 100   & 104           & 7.7        & $-$120.1 & $-$254.3 & 138.2       \\
J1024+4920   &\dots & \dots & He      & 4680 $\pm$ 120   & 128           & 6.8        & $-$80.1  & $-$391.8 & 242.1       \\
J1056+5043   &\dots & \dots & He      & 4560 $\pm$ 70    & 104           & 7.0        & $-$36.7  & $-$119.7 & 61.7        \\
J1101+4854   &\dots & \dots & He      & 4770 $\pm$ 80    & 97            & 6.6        & 73.9   & $-$196.1 & 96.8        \\
J1145+5638   &\dots & \dots & He      & 4780 $\pm$ 120   & 148           & 6.5        & $-$127   & $-$114.3 & 119.8       \\
J1205+5502   &\dots & \dots & He      & 4560 $\pm$ 70    & 102           & 7.0        & $-$23.4  & $-$311.8 & 151.2       \\
J1323+0223   &\dots & \dots & H       & 4250 $\pm$ 170   & 116           & 8.4        & $-$112.2 & $-$50.6  & 67.5        \\
J1333+4942   &\dots & \dots & He      & 4550 $\pm$ 60    & 95            & 7.0        & $-$174.7 & $-$0.9   & 79.0        \\
J1409$-$0100 &\dots & \dots & H       & 4090 $\pm$ 120   & 92            & 8.8        & $-$125.2 & 75.8   & 63.7        \\
J1421+0356   &\dots & \dots & He      & 4340 $\pm$ 100   & 111           & 7.5        & $-$156.2 & $-$15    & 82.3        \\
J1434+5345   &\dots & \dots & He      & 4600 $\pm$ 70    & 100           & 7.0        & $-$143   & 136.6  & 93.5        \\
J1444+6025   &\dots & \dots & He      & 4730 $\pm$ 100   & 132           & 6.6        & $-$18.1  & $-$134.3 & 85.1        \\
J1446+0258   &\dots & \dots & He      & 4360 $\pm$ 140   & 104           & 7.5        & $-$196.2 & 43.2   & 99.5        \\
J1509+5408   &\dots & \dots & He      & 4320 $\pm$ 110   & 114           & 7.6        & 13.3   & $-$135.7 & 73.9        \\
J1513+5023   &\dots & \dots & -3.22   & 3860 $\pm$ 180   & 86            & 8.7        & $-$98.6  & $-$49.4  & 45.0        \\
J1513+4743   & DA   & 1 &  H          & 5960 $\pm$ 120   & 124           & 2.3        & $-$500.8 & $-$147.1 & 305.9       \\
J1515+5930   &\dots & \dots & H       & 3700 $\pm$ 120   & 87            & 9.8        & $-$74.6  & 90.4   & 48.5        \\
J1533$-$0012 &\dots & \dots & He      & 4240 $\pm$ 100   & 107           & 7.8        & $-$44.2  & $-$140.2 & 74.7        \\
J1534+5624   & DC   & 1 & H           & 4900 $\pm$ 120   & 84            & 6.2        & $-$140.3 & 119.5  & 73.2        \\
             &      &   & He          & 5050$_{-70}^{+120}$ & 93        & 5.7        &        &        & 81.0        \\
J1552+4638   & DA   & 1 & H           & 5100$_{-100}^{+120}$ & 88            & 5.2        & $-$48.7  & $-$181.5 & 78.1   \\
J1555+4940   & DC   & 1 & -3.16       & 3910 $\pm$ 180   & 94            & 8.5        & 27.1   & $-$127.1 & 57.6        \\
J1601+4120   & DC   & 1  & H          & 4080 $\pm$ 120   & 36            & 8.8        & 74.8   & $-$228.7 & 40.6        \\
             &      &    & He         & 4610 $\pm$ 60    & 44            & 6.9        &        &        & 50.5        \\
J1601+4204   &\dots & \dots & He      & 4540 $\pm$ 110   & 119           & 7.1        & $-$111.1 & 70.9   & 74.1        \\
J1604+3923   & DA   & 1 & H           & 5010 $\pm$ 140   & 99            & 5.6        & 15.3   & $-$152.1 & 72.0        \\
J1624+4156   & DA   & 1 & H           & 5840 $\pm$ 150   & 133           & 2.4        & 27.7   & $-$194.1 & 123.6       \\
J1627+3726   & DC   & 3 & He          & 4650 $\pm$ 50    & 57            & 6.8        & $-$24.1  & $-$169.3 & 46.0        \\
J1643+4438   & DC   & 1 & He          & $4910 _{-40}^{+60}$ & 70            & 6.2        & 42.8   & $-$197.7 & 66.8      \\
J1647+3946   & DC   & 1 & He          & 4360 $\pm$ 70    & 91            & 7.5        & $-$114.3 & $-$111.8 & 69.3        \\
J1649+2932   & DC   & 1 & He          & 4930 $\pm$ 80    & 99            & 6.2        & 121.3  & 16.1   & 57.6        \\
J1657+2638   & DC   & 1 & H           & 3550 $\pm$ 100   & 67            & 10.1       & $-$73.3  & $-$104.6 & 40.3        \\
J1711+2940   & DC   & 1 & He          & 4550 $\pm$ 70    & 97            & 7.1        & 57.7   & $-$165.8 & 80.7        \\
J1715+2600   & DC   & 1 & He          & 4310 $\pm$ 80    & 88            & 7.6        & $-$34.3  & $-$162.2 & 69.5        \\
J2127+1036   & DC   & 1 & H           & 3970 $\pm$ 370   & 93            & 9.1        & $-$112.5 & $-$65.9  & 57.5        \\
             &      &      & He       & 4470 $\pm$ 160   & 113           & 7.2        &        &        & 69.7        \\
J2230+1415   & DA   & 1 & H           & 5210 $\pm$ 140   & 107           & 4.6        & $-$18.3  & $-$142.3 & 72.7 \\
J2316$-$1044 & DC   & 1  & H          & 3670 $\pm$ 790   & 93            & 9.8        & 237.7  & $-$72    & 109.7       \\
             &      &      & He       & 4310 $\pm$ 200   & 115           & 7.6        &        &        & 136.0       \\
J2320$-$0845 & DC   & 1  & H          & 3240 $\pm$ 190   & 67            & 10.8       & 166.7  & 20.3   & 53.3        \\
             &      &      & He       & 4010 $\pm$ 130   & 80            & 8.3        &        &        & 63.7    \\
\hline
\end{tabular}
\end{table*}

\subsection{Pure H Solutions}

\begin{figure*}
\includegraphics[angle=-90,width = 14cm]{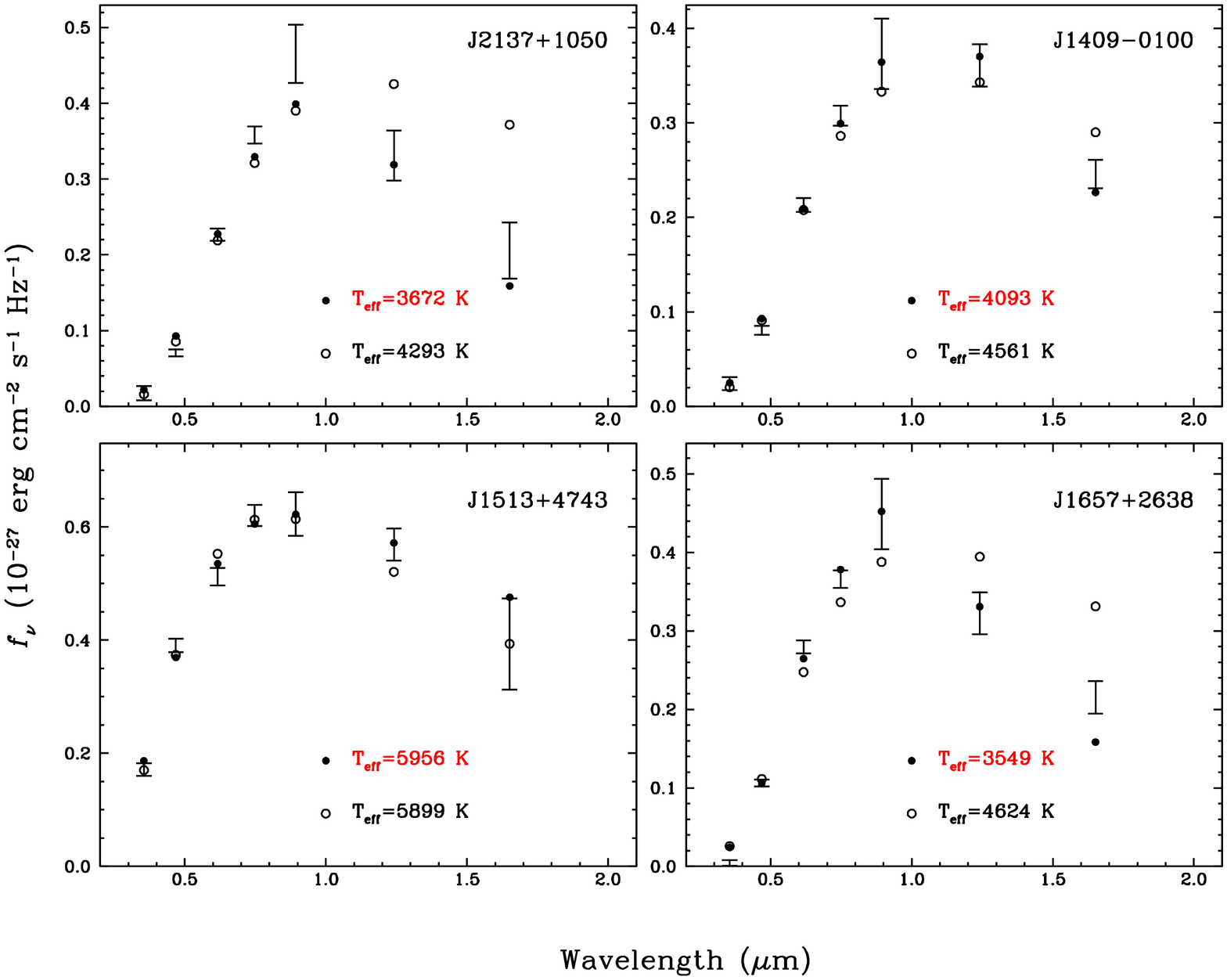}
 \caption{Fits to the SEDs for four of our WDs with pure H atmospheres (full sample available online). Filled circles are pure H models, and open circles are pure He models (included for comparison).}
 \label{fig:pure_h}
\end{figure*}

Of our 57 targets, only 45 have the near-infrared data that are needed to observe the CIA that allows us
to detect the presence of hydrogen. Of these 45 objects, twelve have SEDs best fit by pure hydrogen models.
Figure \ref{fig:pure_h} shows the SEDs and our model fits for four of these objects (full sample is available online). We show the photometric data as error bars and the best-fit model 
fluxes for pure H and pure He composition as filled and open circles, respectively.

J1513+4743 is the only DA white dwarf in our sample with near-infrared photometry available (the four
other DAs are discussed in Section \ref{sec:nonIR}), and the pure H model is a better fit to the SED
than the pure He model. For the remaining objects, we chose the composition based on the solution that
best fits the SED. Our sample includes three previously published H-atmosphere DC white dwarfs: J2137+1050,
J2145+1106N, and J2145+1106S \citep{kilic10a}. Our temperature estimates of $3670\pm 160$, $3720\pm110$, and
$3960\pm100$ K, respectively agree with the previously published values of 3780, 3730 K, and 4110 K \citep{kilic10a}
within the errors. 

With the exception of the DA WD J1513+4743, all of the remaining 11 objects that are best explained by pure H
atmosphere models have $T_{\rm eff} \leq 4250$ K. These objects appear significantly fainter in the $H-$band
than expected from the blackbody-like SEDs of pure He atmosphere white dwarfs, indicating that they have H-rich
atmospheres. In addition to the previously published J2137+1050 and J2145+1106N \citep{kilic10a}, we identify
three new white dwarfs with $T_{\rm eff}\leq3700$ K; namely J0734+3728, J1515+5930, and J1657+2638.
The latter is the coolest white dwarf known ($T_{\rm eff} = 3550 \pm 100$ K) with an SED that is matched
relatively well by a pure H atmosphere model. The implied cooling age for such a cool white dwarf is
10.1 Gyr assuming an average mass, $\log{g}=8$, white dwarf.

\subsection{Pure He Solutions}

\begin{figure*}
 \includegraphics[angle=-90,width = 14cm]{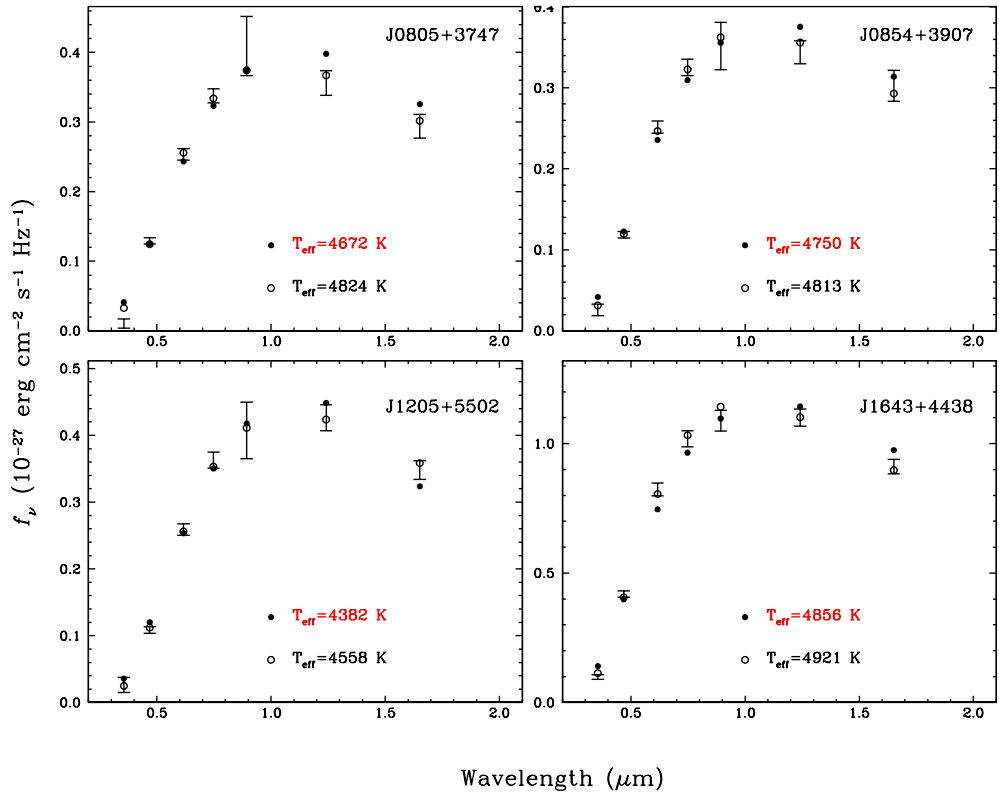}
 \caption{Fits to the SEDs for four of our WDs with pure He atmospheres (full sample available online). Filled circles are pure H models (included for comparison), and open circles are pure He models.}
 \label{fig:pure_he}
\end{figure*}

For our remaining 33 objects with infrared data, 29 show no evidence of CIA and are best fit by
pure He atmosphere models. Figure \ref{fig:pure_he} shows the SEDs for a sample of these targets. All 29 of these
objects have $T_{\rm eff}$ in the range $4240-4930$ K. Nine stars have optical spectra available, and
all nine are DC white dwarfs. 

The differences between the pure H and pure He
model fits are relatively small in this temperature range, and additional $K-$band photometry would be
useful to confirm the atmospheric composition for these stars. However, \citet{bergeron01a} and \citet{kilic10b}
also find an overabundance of pure He atmosphere white dwarfs in the temperature range 4500-5000 K.
\citet{kilic10b} discuss a few potential problems that could lead to misclassification of spectral types
for these stars, including problems with the CIA calculations, or small shifts in the $ugriz$ or $JH$
photometric calibration.

\subsection{Mixed Atmosphere Solutions}

The last four targets with infrared data (J0910+3744, J0927+4852, J1513+4743, and J1555+4940) have SEDs
that are inconsistent with either a pure H or pure He atmosphere solution. We fit the SEDs of these stars
with a mixed H/He atmosphere model. The mixed models allow for significant $H_2$-He CIA at higher temperatures
than seen for $H_2$-$H_2$, as CIA becomes an effective opacity source at higher temperatures in cool He-rich
white dwarfs due to lower opacities and higher atmospheric pressures \citep{bergeron02}. 

\begin{figure*}
 \includegraphics[width = 8cm]{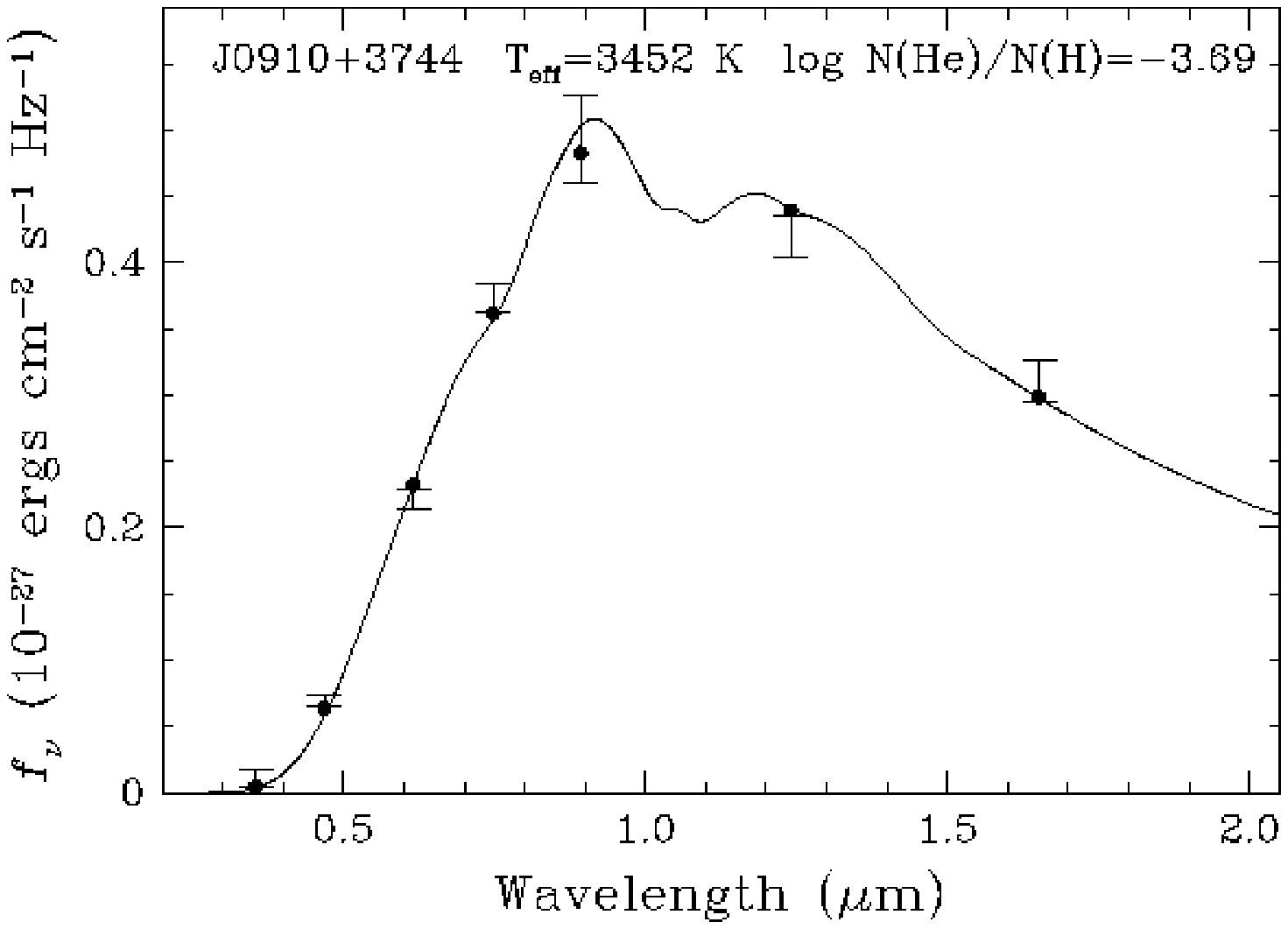}
 \includegraphics[width = 8cm]{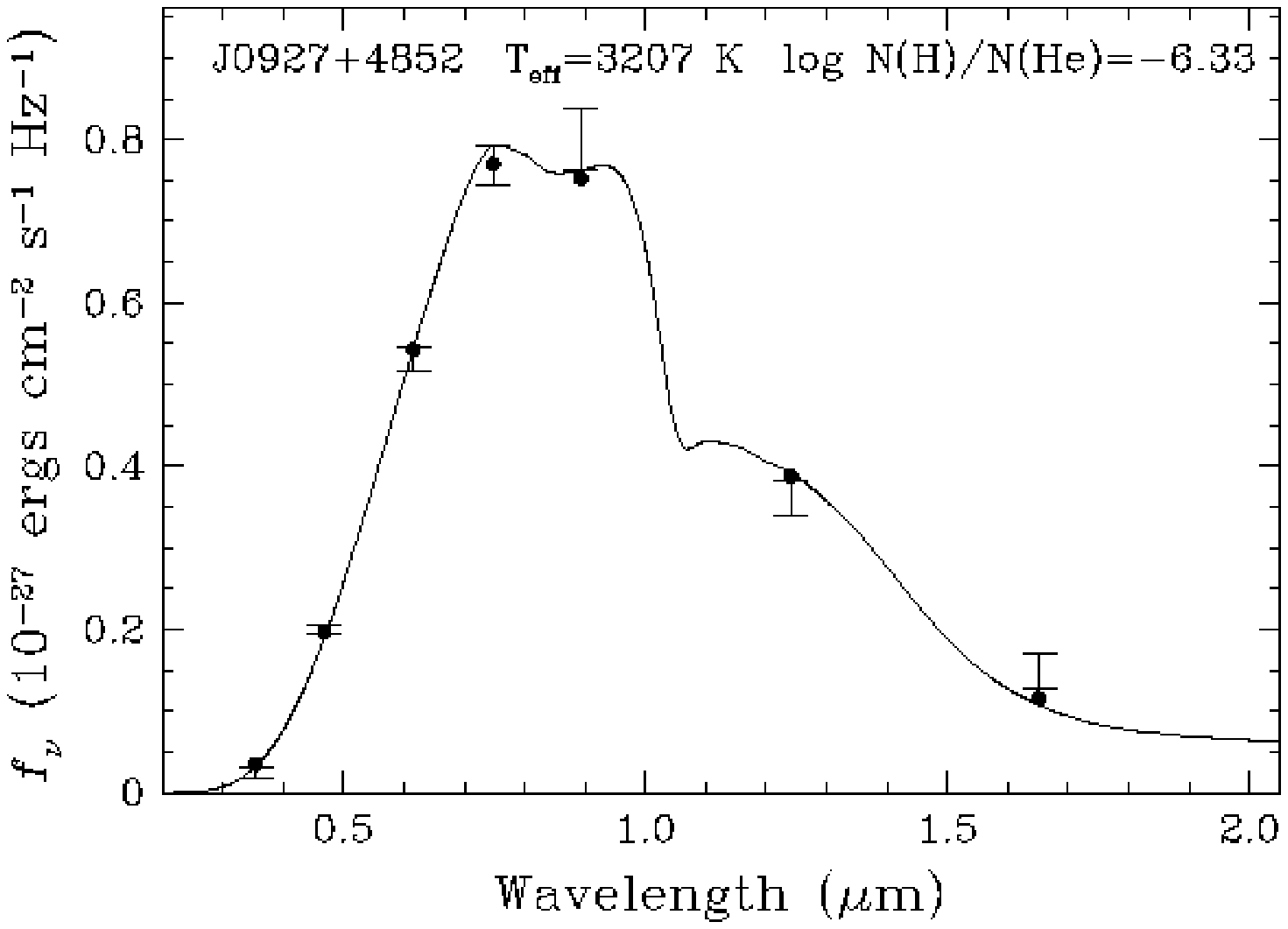}
 \includegraphics[width = 8cm]{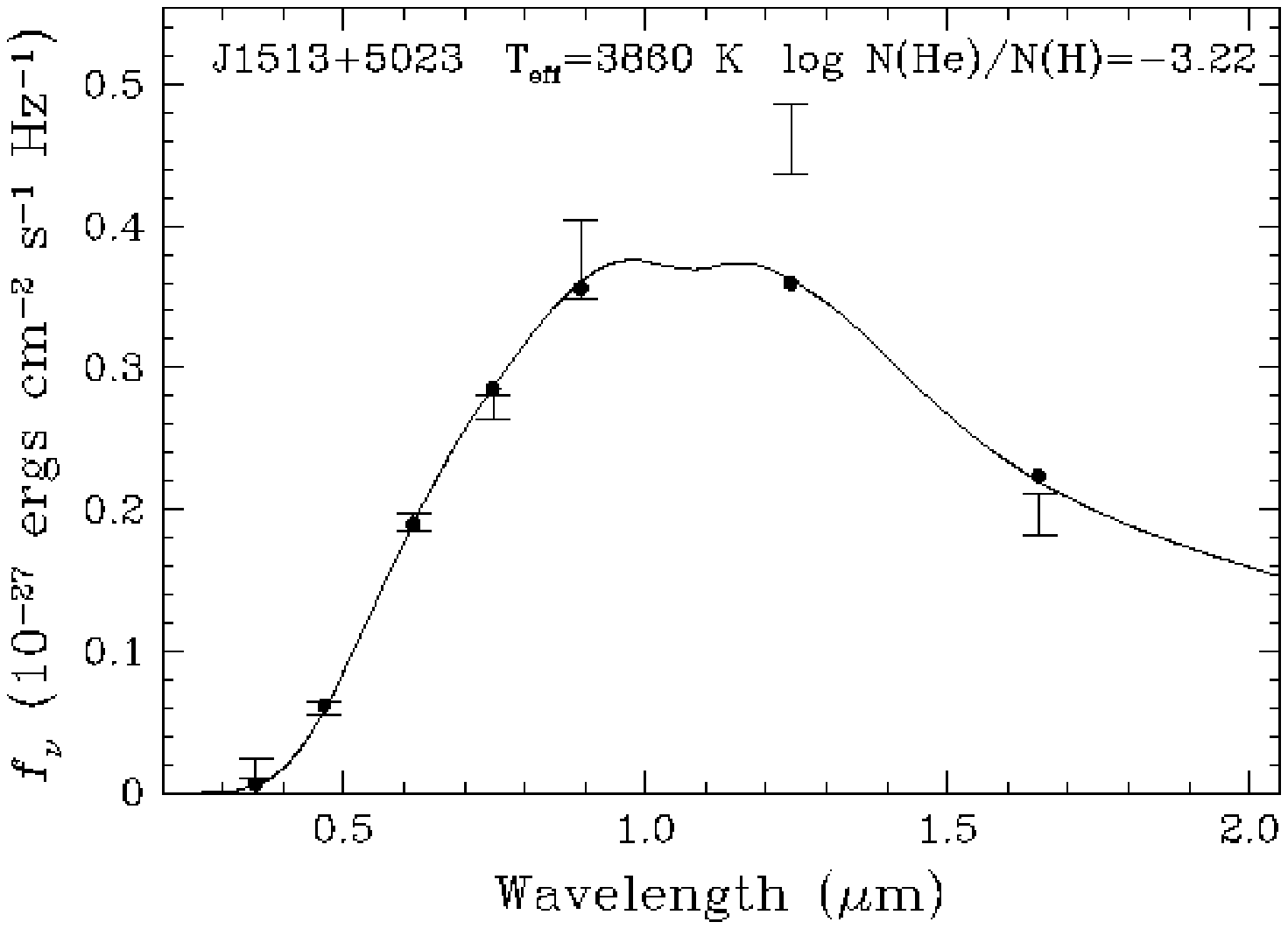}
 \includegraphics[width = 8cm]{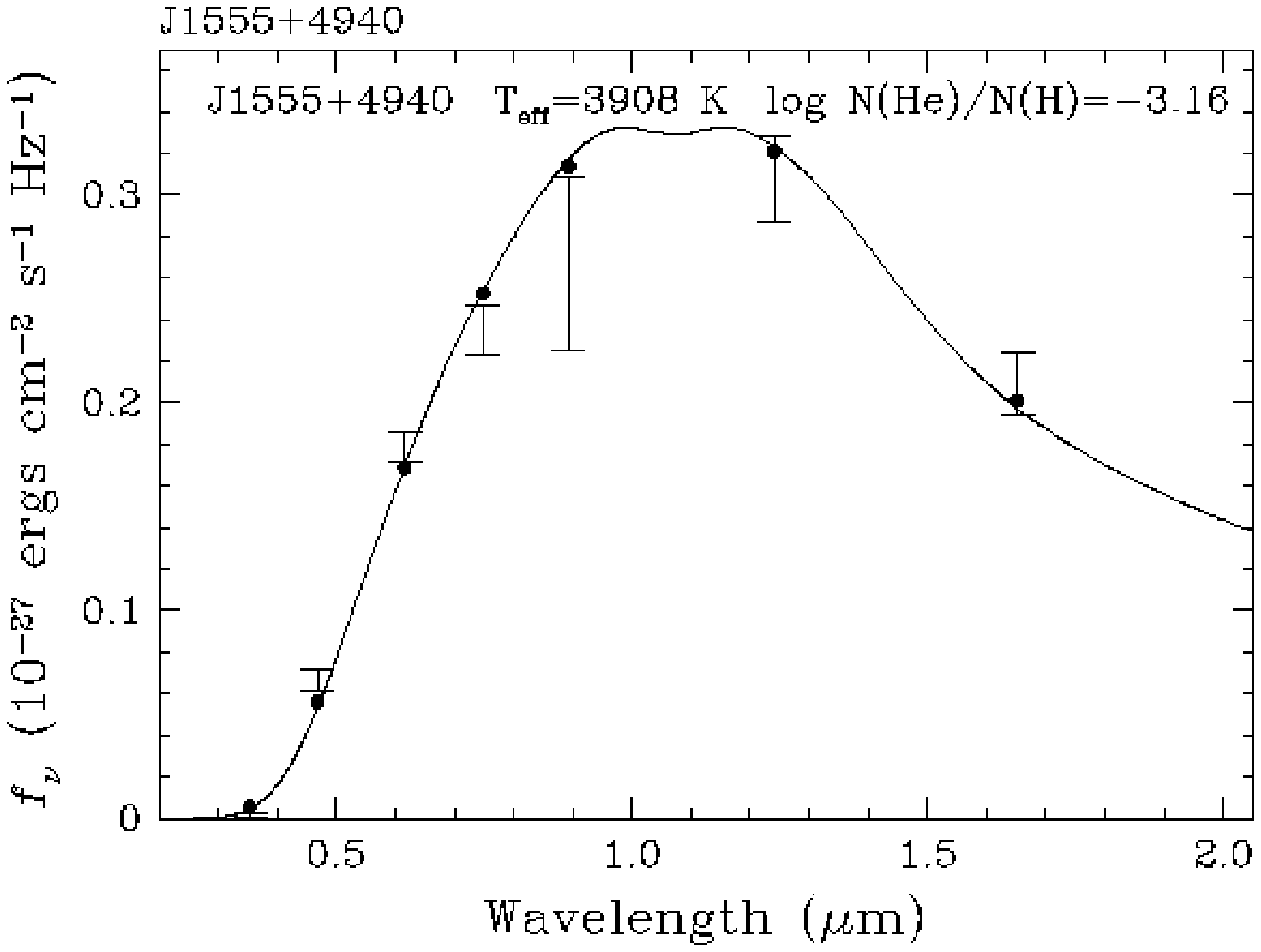}
 \caption{Fits to the SEDs of the four white dwarfs best fit by mixed atmosphere models.}
 \label{fig:mixed}
\end{figure*}

Figure \ref{fig:mixed} shows our mixed H/He atmosphere model fits for these four objects. 
The models yield $\log{(He/H)}$ of $-3.7, 6.3, -3.2$, and $-3.2$ and temperatures of 3450, 3210, 3860,
and 3910 K, respectively. Note that these models predict CIA absorption features around 0.8 and 1.1$\mu$m
that are never observed in cool white dwarfs. Hence, the temperature and composition estimates for such
infrared-faint stars is problematic \citep[see the discussion in][]{kilic10b,gianninas15}. 

The coolest object among these four stars, J0927+4852 appears to be similar to WD0346+246.
\citet{oppenheimer01} originally found a $T_{\rm eff} = 3750$ K and $\log{(He/H)}$ = 6.4 for
WD0346+246, for an assumed surface gravity of $\log{g}=8$. However, \citet{bergeron01a} showed that such a
He-rich atmosphere would require accretion rates from the interstellar medium too low to be
realistic. With the addition of parallax observations to constrain the distance, they estimated
a more realistic solution with $T_{\rm eff} = 3780$ K, $\log{(He/H)}=1.3$, and $\log{g}=8.34$.
A re-analysis by \citet{kilic12} that include the red wing of the Ly$\alpha$ opacity indicate
a similar solution with $T_{\rm eff} = 3650$ K, $\log{(He/H)}=-0.4$, and $\log{g}=8.3$.
Adopting a similar $\log{g}$ value for J0927+4852 would yield a $T_{\rm eff}$ of 3730 K and $\log{(He/H)}$ of 0.3.

This exercise shows the problems with constraining the atmospheric composition of ultracool white dwarfs, and
the need for parallax observations to derive accurate parameters for such white dwarfs.
Regardless of these issues, all four mixed atmosphere white dwarfs appear to be ultracool ($T_{\rm eff}<4000$ K),
bringing the total number of ultracool white dwarfs in our sample to ten.

\subsection{Targets without Infrared Data}
\label{sec:nonIR}

\begin{figure*}
 \includegraphics[angle=-90,width = 14cm]{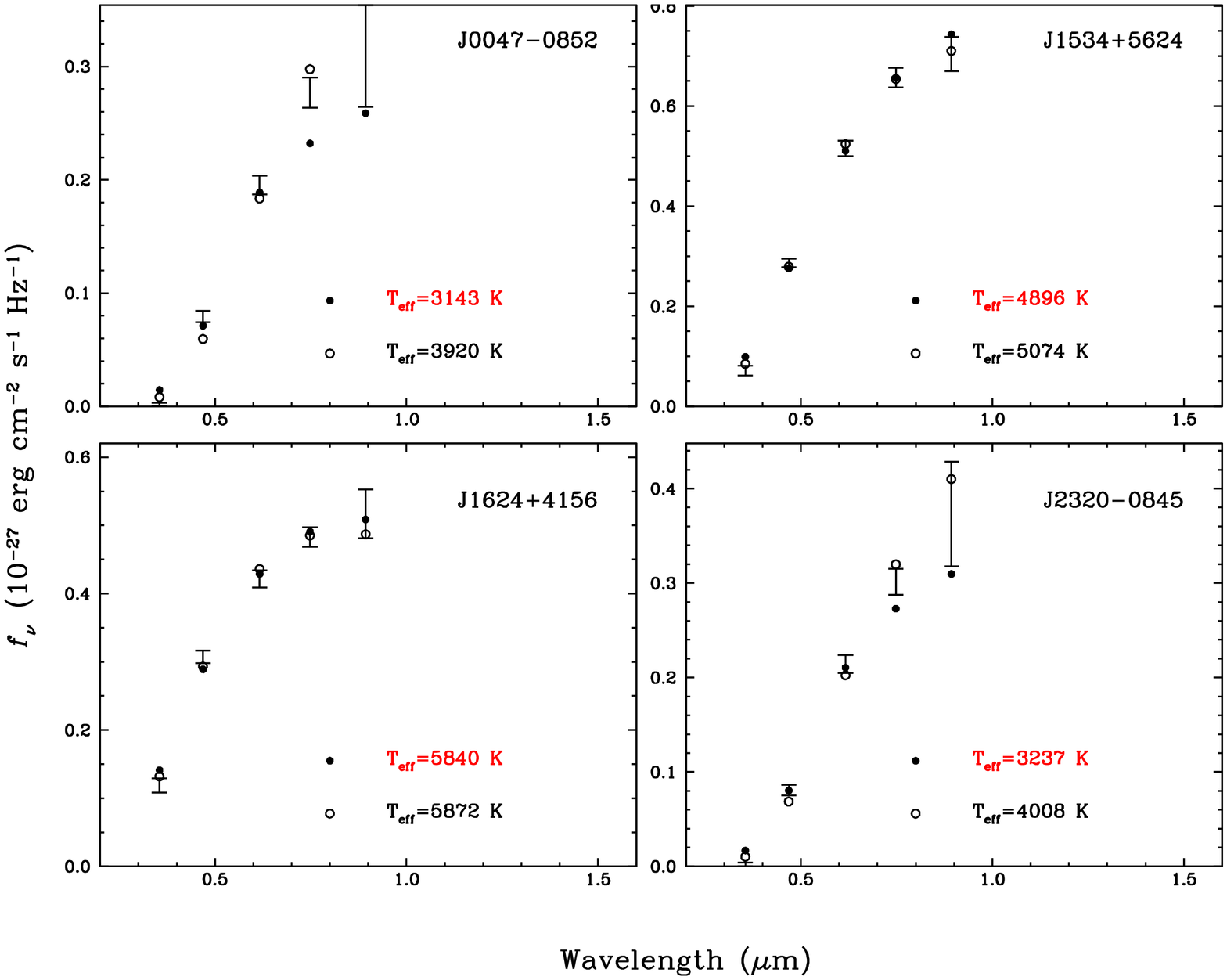}
 \caption{Fits to the SEDs for a sample of our WD lacking IR data (full sample available online). Filled circles are pure H models, and open circles are pure He models. As can be seen, no model is clearly better.}
 \label{fig:non-ir}
\end{figure*}

There are twelve spectroscopically confirmed white dwarfs in our sample that lack infrared photometry.
Figure \ref{fig:non-ir} displays the SEDs along with the pure H and pure He model fits for a subsample of these objects.
The spectra of four of these objects; J1552+4638, J1604+3923, J1624+4156, and J2230+1415 confirm that they
are DA white dwarfs, and the pure H models reproduce the SEDs and spectra reasonably well. This brings our
final number of pure H solutions to sixteen stars.

The remaining eight objects without infrared data are confirmed to be DC white dwarfs. For the most part, the 
pure H and pure He models are nearly indistinguishable in the optical for these objects and we cannot 
determine their composition. Table \ref{tab:wd} shows the results for both pure H and pure He solutions
for these objects. 

All of these targets have SEDs rising toward 1$\mu$m, hence the lack of infrared data
limits the precision of these temperature measurements. However, given the lack of He
atmosphere white dwarfs below 4240 K, we do not expect J0047$-$0852 and J2320$-$0845 to
have pure He atmospheres, and if they were to have pure H atmospheres, they would be the
coolest white dwarfs in our sample, with $T_{\rm eff}$ of $3140 \pm 160$ K and $3240 \pm 190$
K respectively. J0108$-$0954, J2127+1036, and J2316$-$1044 are also potentially ultracool objects
if the pure H solution is correct, which would bring the total number of ultracool white dwarf
candidates in our sample to fifteen. Without infrared data, however, we cannot rule out
the pure He solution, or the possibility of a mixed H/He atmosphere for J0047$-$0852 and J2320$-$0845. 

\section{Kinematic Membership}
\label{sec:kinematics}

The estimated temperatures for our targets yield white dwarf cooling ages between 5 and 10 Gyr,
with the only notable exceptions being J1513+4743 and J1624+4156, which have cooling ages of 2.3
and 2.4 Gyr respectively. Eight objects have cooling ages longer than 9 Gyr, with the oldest
being J1657+2638 at 10.1 Gyr. However, in order to associate a white dwarf with the thick disc or
halo, it is important to determine the total stellar age \citep{bergeron05}. The main-sequence
lifetime of the $\approx 2 M_{\odot}$ progenitor of a $0.6 M_{\odot}$ white dwarf is 1.0-1.3 Gyr;
therefore, the total ages of our objects on average range from 6 to 11 Gyr, with J1513+4743 and
J1624+4156 having total ages between 3.3 and 3.7 Gyr. 

\begin{figure}
 \includegraphics[scale=0.425,bb=30 167 592 679]{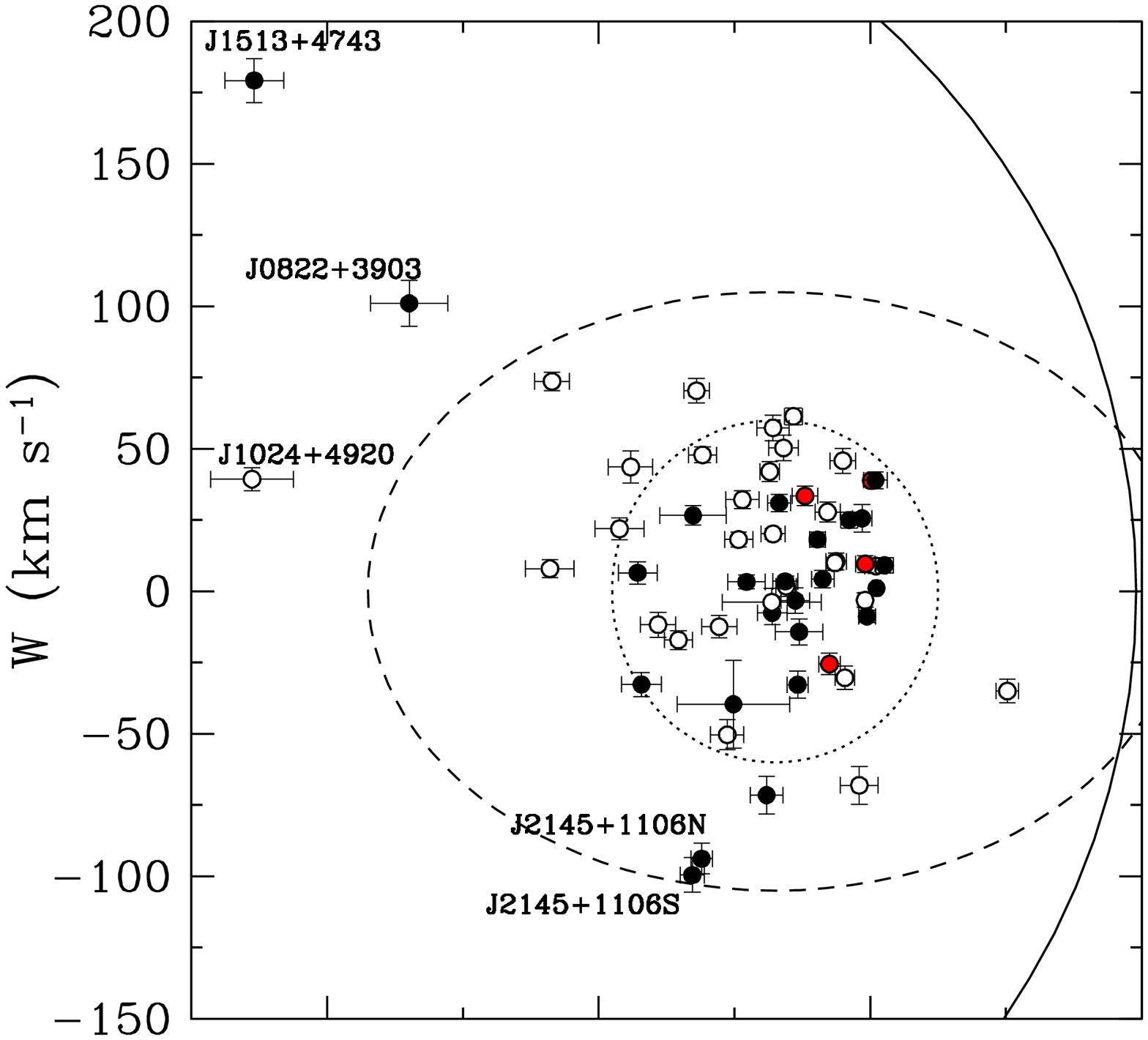}
 \includegraphics[scale=0.425,bb=30 117 592 679]{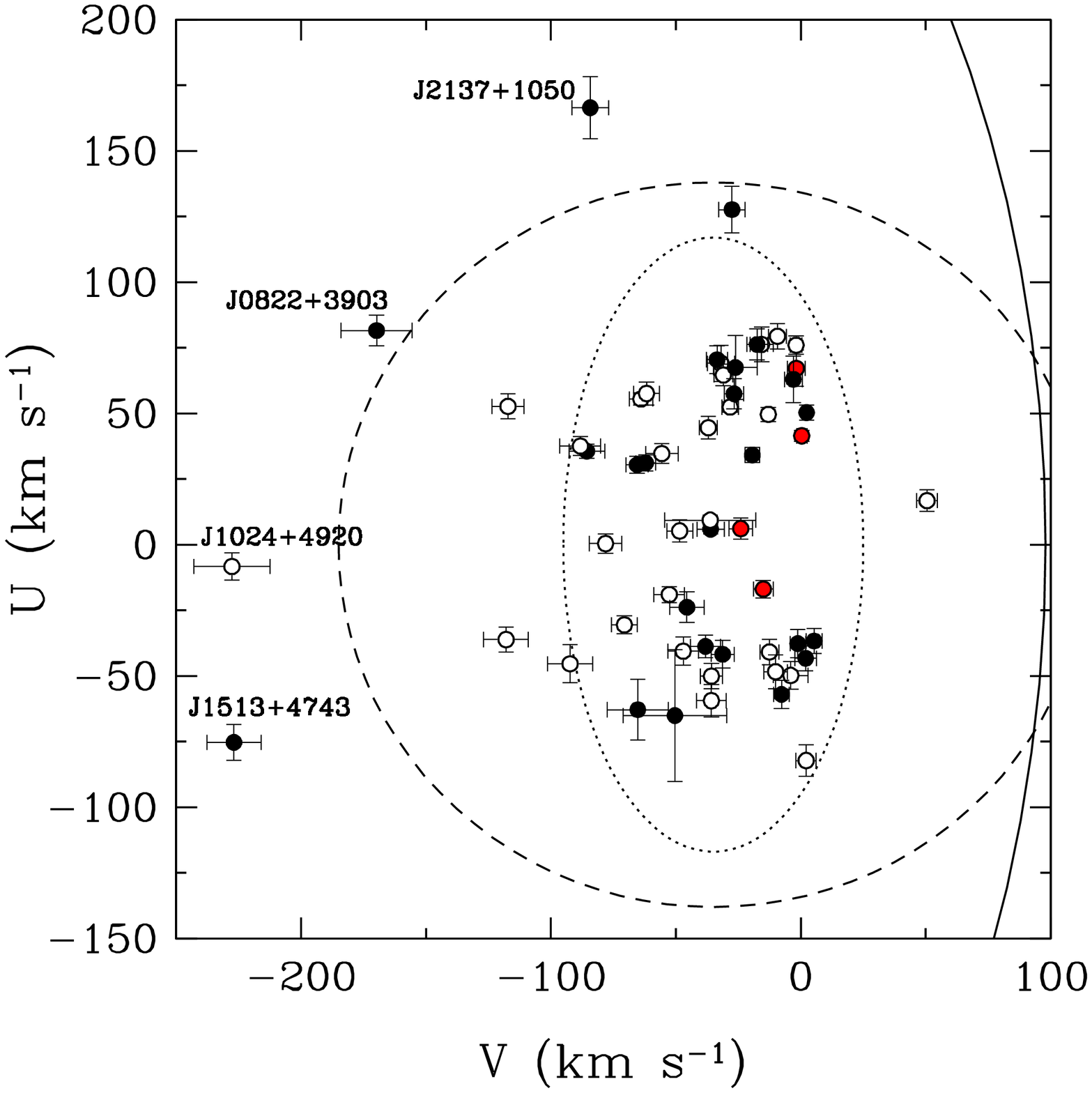}
 \caption{Plots of W vs. V (top) and U vs. V (bottom) velocity distributions for our sample of H-rich (black dots), He-rich (white dots), and mixed (red) WDs. Also plotted are the 3$\sigma$ ellipsoids for the Galactic thin disc (dotted), thick disc (dashed), and stellar halo populations (solid).}
 \label{fig:UVW}
\end{figure}

Figure \ref{fig:UVW} shows U versus V (bottom) and W versus V (top) velocities of our objects
(assuming a radial velocity of 0 km s$^{-1}$and calculated using the prescription of \citet{johnson87}), 
as well as the 3$\sigma$ ellipsoids of the halo, thick disc, and thin disc populations \citep{chiba00}. 
The filled, open, and red circles represent the objects best fit by pure H, pure He, and mixed H/He 
atmosphere models, respectively. For the eight objects with undetermined compositions, velocities were 
calculated assuming the pure H solution for simplicity. The choice of the pure H or pure He solution 
has a negligible effect on the final UVW velocities (see Table \ref{tab:wd}). 

\begin{figure*}
 \includegraphics[angle=-90,width = 14cm]{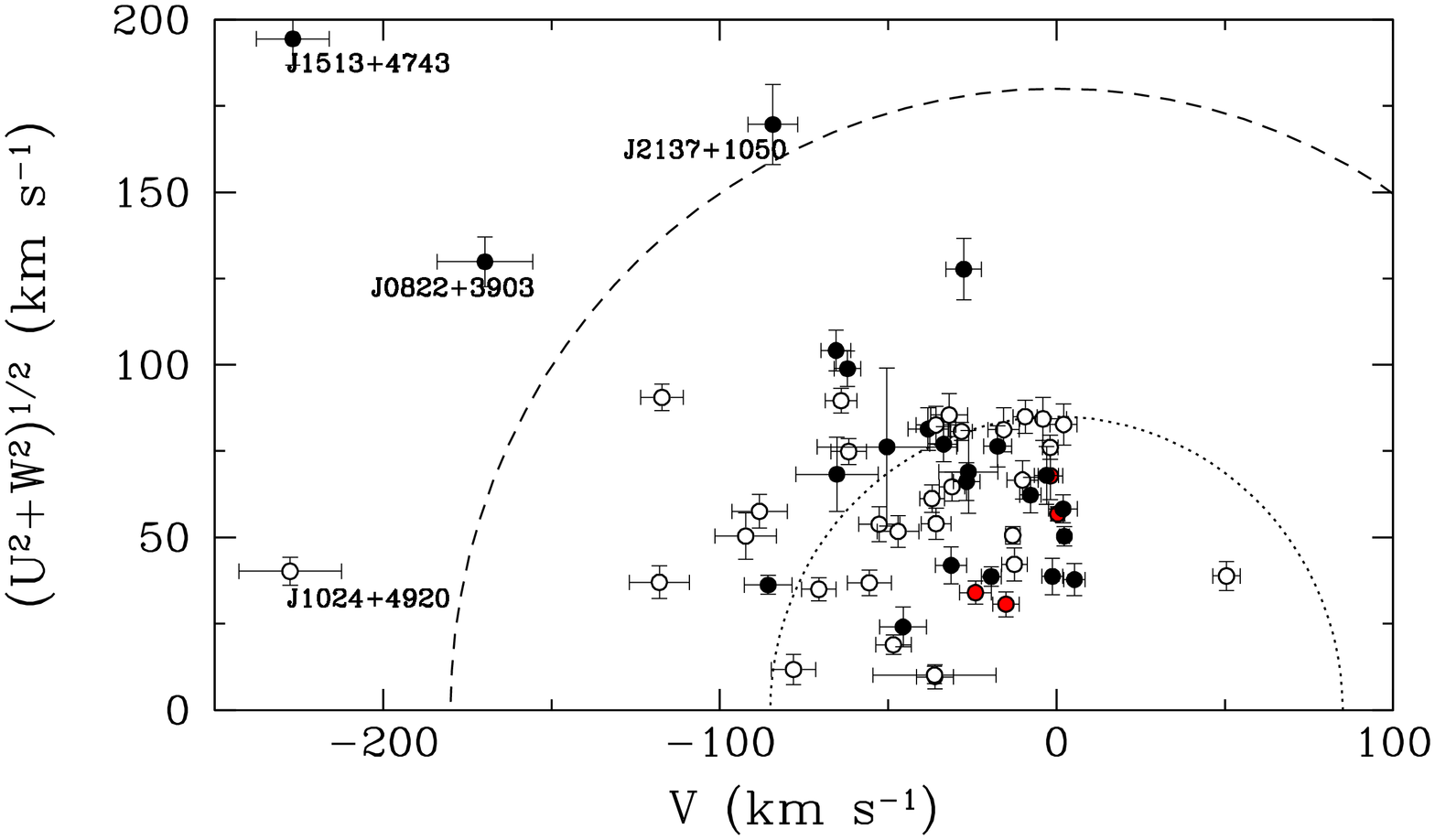}
 \caption{Toomre diagram for our 57 targets. Symbols have the same meaning as Figure \ref{fig:UVW}. Thin disc (dotted) and thick disc (dashed) boundaries taken from \citet{fuhrmann04}.}
 \label{fig:toomre}
\end{figure*}

J2137+1050 shows velocities inconsistent with thick disc objects in U, consistent with the analysis in \citet{kilic10a}, 
while the results for the J2145+1106 common-proper motion binary are consistent to 2$\sigma$, but not 3$\sigma$. In addition, 
three other targets in our sample show velocities inconsistent with thick disc objects: J0822+3903, J1024+4920, and
J1513+4743, with  cooling ages of 8.5, 6.8, and 2.3 Gyr respectively. The Toomre diagram for our targets is shown in 
Figure \ref{fig:toomre}, with thin disc and thick disc boundaries from \citet{fuhrmann04}; the differentiation between our halo 
candidates and the rest of our sample is clearer here than in Figure \ref{fig:UVW}.   

The total main-sequence + white dwarf cooling ages of these objects are relatively young for halo objects,
but without parallax measurements, we cannot constrain their masses, velocities, and cooling ages precisely. For example,
if these objects have $M\approx0.53M_{\odot}$ \citep{bergeron01a}, the progenitor mass would be closer to $1M_{\odot}$ and 
their main-sequence lifetimes would be on the order of 10 Gyr, making them excellent candidates for membership in the halo. 
A lower surface gravity would also imply a larger and more distant white dwarf, and UVW velocities that are even more inconsistent 
with the thick disc population. Conversely, for $\log{g} = 8.5$ white dwarfs, the cooling ages would range from 5 to 11
Gyr, and the UVW velocities of our halo white dwarf candidates would remain inconsistent with thick disc objects. 

Interestingly, with the exception of the three previously published white dwarfs (J2137+1050 and J2145+1106 binary),
none of our objects with cooling ages above 9 Gyr have UVW velocities inconsistent with the thick disc, nor do they 
show the high tangential velocities expected for halo objects. In fact, the highest tangential velocity for these 
objects is 72 km s$^{-1}$. Assuming these objects really do belong to the thick disc gives a thick disc age of 
$\approx$11 Gyr.

Our assumption of zero radial velocity has a negligible effect on our results \citep[see the discussion in][]{kilic10a}.
The UVW velocities of our halo white dwarf candidates remain inconsistent with the 3$\sigma$ distribution for the thick 
disc for positive and negative radial velocities up to 100 \kms (though J0822+3903 only remains inconsistent in both U and
W for radial velocities between -90 and 30 \kms).

\section{Conclusions}
\label{sec:con}

We present follow-up optical spectroscopy and/or near-infrared photometry of 57 cool white dwarf candidates
identified from a $\approx3100$ square degree proper motion survey described by \citet{munn14}. Thirty one
of our candidates are spectroscopically confirmed to be white dwarfs, including 5 DA and 26 DC white dwarfs.
The remaining targets have proper
motion measurements from both optical and infrared observations that are consistent within the errors. The
optical and near-infrared colors for these targets are also consistent with the predictions from the white
dwarf model atmospheres. Hence, the contamination from subdwarfs should be negligible for this sample of
57 stars.

We perform a model atmosphere analysis of these 57 objects using $ugriz$ and $JH$ photometry. 
The best-fit models have 29 pure He atmosphere white dwarfs with $T_{\rm eff}=4240-4930$ K, 16 pure
H atmospheres with $T_{\rm eff}=3550-5960$ K, and 4 mixed H/He atmospheres with $T_{\rm eff}=3210-3910$ K.
Eight of our targets lack the near-infrared data necessary to differentiate between the pure H and pure He solutions.

Our sample contains ten ultracool white dwarf candidates, with another five potential candidates that currently
lack near-infrared data. All of the ultracool white dwarfs have hydrogen-rich atmospheres. J1657+2638 is
the most interesting with $T_{\rm eff} = 3550 \pm 100$ K and an SED that is reproduced fairly well by
a pure H atmosphere. For an average mass of 0.6 $M_{\odot}$, J1657+2638 would be an $\approx$11 Gyr old
(main-sequence + cooling age) white dwarf at a distance of 67 pc. The implied tangential velocity of 40 km s$^{-1}$
demonstrates that J1657+2638 belongs to the Galactic thick disc.

Our sample contains three new halo white dwarf candidates. All three have high tangential velocities and
UVW velocities inconsistent with the Galactic thick disc. The oldest halo white dwarf candidate
is J0822+3903 with a cooling age of 8.5 Gyr. However, without trigonometric parallax observations, we cannot
accurately constrain the distances, masses, and ages of our white dwarfs.  

Our current sample of cool field halo white dwarfs is limited by a lack of deep proper motion surveys. Ongoing
and future large scale surveys such as \textit{GAIA} and \textit{LSST} will find a significant number of cool
white dwarfs, including halo white dwarfs, in the solar neighborhood.  With $g-$band magnitudes of 20$-$22, we 
expect parallax errors from \textit{GAIA} to range from about 400$-$1200 $\mu$as 
\footnote{http://www.cosmos.esa.int/web/gaia/science-performance}, corresponding to uncertainties of $\approx 20$ 
per cent in both mass and cooling age for the majority of our targets. In addition, \textit{GAIA} will reveal the 
brighter population of halo white dwarfs near the Sun. 

\section*{Acknowledgements}

We gratefully acknowledge the support of the NSF and
NASA under grants AST-1312678 and NNX14AF65G, respectively.

\end{document}